\documentclass[reprint,aps,prd,nofootinbib,a4paper,10pt,twocolumn,superscriptaddress]{revtex4-2} %

\usepackage{amsmath,amssymb,graphicx}
\usepackage{appendix,bm,dcolumn}
\usepackage[T1]{fontenc}
\usepackage[utf8]{inputenc}
\usepackage{color}
\usepackage[unicode=true,pdfusetitle,bookmarks=true,bookmarksnumbered=false,bookmarksopen=false,breaklinks=true,pdfborder={0 0 0},backref=false,colorlinks=true]{hyperref}
\hypersetup{citecolor=blue,filecolor=blue,linkcolor=blue,urlcolor=blue}
\usepackage{pstool}
\usepackage{subcaption}
\usepackage{multirow,aas_macros}

\newcommand{\sayy}[1]{`#1'}
\def\Eu{ \mathfrak{H} } 
\def\la{\langle}
\def\ra{\rangle}
\def\bea{\begin{eqnarray}}
\def\eea{\end{eqnarray}}

\newcommand{\lcdm}{$\Lambda {\rm CDM}$}
\newcommand{\flrwsolver}{\texttt{FLRWSolver}}
\newcommand{\mesc}{\texttt{mescaline}}
\newcommand{\alp}{\alpha}
\newcommand{\gam}{\gamma}
\newcommand{\Gam}{\Gamma}
\newcommand{\pd}{\partial}
\newcommand{\dtwoh}{\frac{{\rm d}^2\mathfrak{H}}{{\rm d}\lambda^2}}
\newcommand{\zeff}{z_{\rm eff}}

\begin{document}

\preprint{APS/123-QED}

%\hypersetup{citecolor=blue,filecolor=blue,linkcolor=blue,urlcolor=blue}

%\shorttitle{The importance of anisotropy in sky-sampling}
%\shortauthors{Macpherson \& Heinesen}

% Working title... change however you wish!!
\title{Luminosity distance and anisotropic sky-sampling at low redshifts: a numerical relativity study}

%% Note that the corresponding author command and emails has to come
%% before everything else. Also place all the emails in the \email
%% command instead of using multiple \email calls.

%\correspondingauthor{Hayley J. Macpherson}

\author{Hayley J. Macpherson}
\email{h.macpherson@damtp.cam.ac.uk}
\affiliation{Department of Applied Mathematics and Theoretical Physics, Cambridge CB3 0WA, UK}

\author{Asta Heinesen}
\email{asta.heinesen@ens-lyon.fr}
\affiliation{Univ Lyon, Ens de Lyon, Univ Lyon1, CNRS, Centre de Recherche Astrophysique de Lyon UMR5574, F--69007, Lyon, France}

\date{\today}% It is always \today, today,
             %  but any date may be explicitly specified

\begin{abstract}
Most cosmological data analysis today relies on the Friedmann-Lema\^{\i}tre-Robertson-Walker (FLRW) metric, providing the basis of the current standard cosmological model. Within this framework, interesting tensions between our increasingly precise data and theoretical predictions are coming to light. It is therefore reasonable to explore the potential for cosmological analysis outside of the exact FLRW cosmological framework. In this work we adopt the general luminosity-distance series expansion in redshift with \emph{no assumptions} of homogeneity or isotropy. This framework will allow for a full model-independent analysis of near-future low-redshift cosmological surveys. 
We calculate the effective observational \sayy{Hubble}, \sayy{deceleration}, \sayy{curvature} and \sayy{jerk} parameters of the luminosity-distance series expansion in numerical relativity simulations of realistic structure formation, for observers located in different environments and with different levels of sky-coverage. 
%\hmr{update numbers here} 
With a \sayy{fairly-sampled} sky, we find {0.6\% and 4\%}
%2\% and 15\% 
cosmic variance in the \sayy{Hubble} and \sayy{deceleration} parameters {for scales of 200 Mpc/h (corresponding to density contrasts of {$\sim 0.05$} 
%$\sim 0.1$ 
in the simulated model universe)}, respectively. On top of this, we find that typical observers measure maximal sky-variance of {2\% and 120\%}
%7\% and 550\% 
in the same parameters{, as compared to their analogies in the large scale FLRW model.} Our work suggests the inclusion of low-redshift anisotropy in cosmological analysis could be important for drawing correct conclusions about our Universe. 

\end{abstract}

%\keywords{Suggested keywords}%Use showkeys class option if keyword
                              %display desired
\maketitle

%% See the online documentation for the full list of available subject
%% keywords and the rules for their use.
%\keywords{cosmology}

\section{Introduction}

Modern cosmology has been shaped by the success of the standard $\Lambda$ Cold Dark Matter (\lcdm) model in providing a consistent fit to almost all of our cosmological observations to date. At the base of this model lies the assumption that the Universe is everywhere close to a Friedmann-Lema\^{\i}tre-Robertson-Walker (FLRW) solution to the Einstein field equations, in such a way that the FLRW model is an accurate lowest order description of the kinematic and dynamic properties of the Universe. 
Among the observations consistently fit within \lcdm\ are the temperature fluctuations in the cosmic microwave background (CMB) radiation \citep[e.g.][]{PlanckCMB:2020}, the baryon acoustic oscillation (BAO) imprint on the galaxy distribution \citep[e.g.][]{Beutler:2011,Alam:2017}, and the distances to supernovae of type Ia \citep[SNIa; e.g.][]{Riess:2007,Abbott:2019}. It is worth noting that the \lcdm\ model has historically been adjusted by the addition of dark matter and dark energy in order to facilitate the explanation of data. These dark components remain unexplained in the standard model of particle physics. 

Amongst the successes of the \lcdm\ model lie some interesting tensions between our observational data and theoretical predictions. The most prominent tension in the \lcdm\ paradigm is that of the inferred value of the Hubble parameter ($H_0$) from CMB observations and the direct measurement of $H_0$ from observations of nearby SNIa and Cepheids \citep{Riess:2018,Riess:2019cxk,DiValentino:2019qzk,Aghanim:2018eyx}.
While many attempts have suggested mechanisms for a partial or total relief from this tension, no one solution has yet been accepted by the community, and so the search continues \citep[see, e.g.,][]{DiValentino:2021,Bernal:2021,DiValentino:2020a}. 

The assumption of a spatially homogeneous and isotropic FLRW model underpins analytic distance relations used to interpret most cosmological datasets. For example, the luminosity-distance redshift relation of the class of flat FLRW metrics is used for, e.g., the measurement of the Hubble constant using the local distance ladder \citep{Riess:2018,Riess:2019cxk} and the measurement of present epoch acceleration of the Universe \citep{Riess:1998,Perlmutter:1999}.
Similarly, the FLRW geometrical assumption is at the core of conventional detections of the BAO feature in the matter distribution \citep{Cole:2005sx,Eisenstein:2005su} and the inference and analysis of the CMB Planck power spectrum \citep{Aghanim:2019ame,Aghanim:2018eyx}. 

The sparsity of available cosmological data has historically made the FLRW assumption justified, since exact model symmetries were needed in order to \emph{a priori} constrain the model universe enough to infer cosmological information. % from its fit to data. 
However, 
%\hmt{for} near future cosmological surveys 
the amount of data is predicted to grow by orders of magnitude in near future surveys 
%relative to \hmt{presently} available surveys
\citep[e.g.][]{Amendola:2018,Ivezic:2019,SKAWG:2020}. As an example, upcoming datasets of SNIa in the 2020's will count hundreds of thousands of supernovae and cover a large proportion of the sky -- see \cite{Scolnic:2019apa} and references therein. 
%As a consequence, 
%\hmt{The} constraining power of data for determining the relation between luminosity distance and redshift will \hmt{therefore} improve significantly. 
The improved constraining power of datasets like these
will allow current model assumptions 
%\emph{a priori} made in present data analysis 
to be relaxed and open the door for fully model independent analysis of cosmological data. 

%\hmr{I removed the following sentence because we talk about this in the next section}
%The series expansion of luminosity distance in redshift can be written in terms of geometric quantities -- in a so called \sayy{cosmographic} representation -- independent on the field equations governing the scale factor of the FLRW metric \citep{Visser:2003vq}. 
A number of studies have taken steps towards this goal 
%A number of studies have generalised this 
and considered general, analytic  %considered
%the series expansion of 
distance measures %in general spacetimes  
\citep[e.g.,][]{Seitz:1994xf,1966ApJ...143..379K,Clarkson:2011uk,Clarkson:2011br,Heinesen:2020b}.
%In \citet{Heinesen:2020b} a cosmographic representation of the luminosity distance in a \emph{general} space-time was presented, without assumptions on the metric tensor of the Universe or the field equations prescribing it. 
Specifically, \citet{Heinesen:2020b} presented 
%a cosmographic representation of 
the series-expanded luminosity distance redshift relation 
%in a \emph{general} spacetime, 
without making assumptions on the metric tensor of the Universe, or the field equations prescribing it. 
%, with the geometric coefficients computed up till third order in redshift. 
%, where  series expansion of   presented a series expansion of the luminosity-distance to third order in redshift valid for a \emph{general} space-time, with no assumptions about the metric of the Universe or the field equations prescribing it. 
Coefficients in this
%luminosity distance 
series expansion %in redshift %-- computed to third order in \citet{Heinesen:2020b} -- 
contain a \emph{finite} set of physically interpretable geometric degrees of freedom describing the luminosity distance in a given direction on the observer's sky. 
%As detailed in \citet{Heinesen:2020b}, t
This representation 
%of the luminosity distance 
allows for fully model-independent analysis
%of large surveys
of low redshift data. %of standardisable low-redshift objects. %cosmographic representation
%\hmr{I removed the rest of this paragraph, since we talk more about this in the relevant section}The anisotropic parameters which generalise the Hubble, deceleration, jerk and curvature parameters of the FLRW series expansion of luminosity distance are non-trivial \hmt{. For instance,} the effective observational deceleration parameter, which takes the place of the FLRW deceleration parameter, can be negative without any local acceleration of space. % in the series expansion

Until cosmological surveys have reached the size and sky coverage necessary for  model-independent analyses,
%needed in order to carry out a full model independent analysis, \hmr{we said this in the prev sentence, sounds repetitive}
it is useful to investigate the expected impact of inhomogeneity and anisotropy within this framework in realistic numerical simulations. 
%\ahr{I moved the following paragraph, as I think it fits well here:}
Numerical relativity (NR) has proven a promising avenue for cosmological simulations of nonlinear structure formation without assumptions of a global \sayy{background} metric of the Universe \citep{Giblin:2015vwq,Bentivegna:2015flc,Macpherson:2016ict,East:2017qmk,Daverio:2019gql}. %See also, e.g., \citep{Adamek:2016,Barrera-Hinojosa:2019mzo} for general-relativistic N-body simulations with some reasonable approximations. 

%Such simulations have already been used to study, e.g., backreaction effects and differential expansion \citep{Bentivegna:2015flc,Macpherson:2018akp,Macpherson:2019a}, observational signatures in the weak-lensing convergence power spectrum \citep{Giblin:2017ezj}, and the accuracy of Newtonian dynamics \citep{East:2017qmk} and linearised gravity \citep{Giblin:2018ndw}.
%\ahr{Perhaps references to some promising simulation tools (including your own :) ).} %gravity or geometry. 
%Simulations \hmt{like these go} beyond the assumptions underlying the \lcdm\ model, \hmt{and can help us understand and} quantify relativistic effects on upcoming precision observations. % using realistic initial data and matter models. 

In this work, we calculate the coefficients of the general luminosity distance series expansion in realistic cosmological simulations performed with numerical relativity. 
%\ahr{cite Macpherson et al here? :) }\hmr{We're using new simulations here though! Not the same ones as my previous papers} 
We sample structures on scales %of 100 Megaparsec (Mpc) and above, 
where the FLRW assumption is considered valid in most cosmological analysis. Considering observers in different local environments, and with different levels of sky coverage, allows us to assess the impact of both inhomogeneity and anisotropy on effective cosmological parameters. This work is a first step towards an upcoming in-depth analysis on the effect of survey geometry on the interpretation of observations in a locally anisotropic universe.

%In Section~\ref{sec:FLRWdL} we review the series expansion for the luminosity-distance redshift relation in an FLRW space-time. 
In Section~\ref{sec:dL} we review the \emph{general} luminosity-distance redshift relation presented by \citet{Heinesen:2020b}, and used in this paper. In Section~\ref{sec:sims} we present details of our simulations, including initial data, gauge, and post-processing analysis. We present our results in Section~\ref{sec:results} and conclude in Section~\ref{sec:conclude}. Greek indices represent space-time indices and take values $0 \ldots 3$, while Latin indices represent spatial indices and take values $1 \ldots 3$, and repeated indices imply summation. We use geometric units with $G=c=1$. %\ahr{Do we state otherwise anywhere?} 

%\ahr{Suggestion: Merge FLRW and general lumninosity sections into one main section with two subsections?.} \hmr{go ahead and try it out!}\ahr{OK, I did this now!} 

\section{The luminosity distance in a general space-time} \label{sec:dL}
Here we formulate the series expansion of luminosity-distance in redshift, valid in a general universe model.  
We first give a brief review of the FLRW expression for the series expanded luminosity distance in Section~\ref{sec:FLRWdL}, after which we provide the analogous expression valid for a general space-time setting in Section~\ref{sec:GeneraldL}. 

\subsection{FLRW luminosity-distance redshift relation}\label{sec:FLRWdL}

Before describing the luminosity distance in a general geometry, we review the series expansion for luminosity distance in redshift under the FLRW geometrical assumption 
as per, e.g., \citet{Visser:2003vq}. We consider a class of emitters and observers with worldlines that are orthogonal to the homogeneous and isotropic spatial sections of the FLRW model. 
In this setting, the luminosity distance, $d_L$, between a causally connected pair of emitters and observers is determined up to third order in redshift, $z$, by
%\begin{equation}\label{eq:dLFLRW}
%    d_{L,{\rm FLRW}}(z) = \frac{1}{H_o} z + \frac{1 - q_o }{2 H_o} z^2 + \frac{- 1 +  3 q_o^2 + q_o    -  j_o   + \Omega_{ko} }{ 6  H_o} z^3   + \mathcal{O}(z^4) \, ,
%\end{equation} 
\begin{equation}\label{eq:dLFLRW}
\begin{aligned}
\hspace*{-0.15cm}    d_{L,{\rm FLRW}}(z) = d_{L,{\rm FLRW }}^{(1)} z   + d_{L,{\rm FLRW}}^{(2)} z^2 +  d_{L,{\rm FLRW}}^{(3)} z^3 \, . %+ \mathcal{O}( z^4) \, . 
\end{aligned}
\end{equation} 
The coefficients 
\begin{equation}\label{eq:dLFLRWcoef}
\begin{aligned}
    d_{L,{\rm FLRW}}^{(1)} &\equiv \frac{1}{H_o} \, ,  \qquad  d_{L,{\rm FLRW}}^{(2)} \equiv \frac{1 - q_o }{2 H_o}  \, , \\%\qquad  
    d_{L,{\rm FLRW}}^{(3)} &\equiv \frac{- 1 +  3 q_o^2 + q_o    -  j_o   + \Omega_{ko} }{ 6  H_o}    \, 
\end{aligned}
\end{equation} 
of the series expansion are given in terms of the Hubble, deceleration, jerk and curvature parameters: 
\begin{equation}\label{eq:dLFLRWdefs}
\begin{aligned}
    H &\equiv \frac{\dot{a}}{a} \, , \qquad \qquad q \equiv -  \frac{\ddot{a}}{a H^2} \, ,\\% \qquad
    j &\equiv  \frac{\dot{\ddot{a}}}{a H^3} \, , \qquad \Omega_{k} \equiv \frac{-k}{a^2 H^2 }  \, ,
\end{aligned}
\end{equation} 
where $a$ is the scale factor, an overdot represents a derivative with respect to the FLRW proper time,  
%denotes derivation with the FLRW proper time parameter, 
and $k \in \{-1,0,1\}$ determines the spatial sections as being \sayy{open}, \sayy{flat} or \sayy{closed}, respectively. 
The subscript $o$ indicates evaluation at the point of observation. 
No assumptions about the field equations governing the scale factor have been imposed in formulating \eqref{eq:dLFLRW}--\eqref{eq:dLFLRWdefs}. Homogeneous and isotropic cosmological analysis which does not rely on the assumptions on field equations is sometimes referred to as \sayy{FLRW cosmography} \citep{1972gcpa.book.....W}.
The vast majority of analyses of cosmological surveys, for instance of standardisable candles, is based on either FLRW cosmography or FLRW predictions within a particular field theory. An example of the latter is the use of general relativity with a dust source and a cosmological constant, which is used to model the late Universe in the \lcdm\ model. 
As we shall see in the following section, the FLRW parameters (\ref{eq:dLFLRWdefs}), and the corresponding coefficients (\ref{eq:dLFLRWcoef}) of the luminosity distance series expansion, generalise in non-trivial ways in the presence of inhomogeneity and anisotropy, with possibly crucial implications for the interpretation of cosmological data.

\subsection{General luminosity-distance redshift relation}\label{sec:GeneraldL}

We now consider the luminosity distance to astrophysical sources as a function of their redshift in \emph{general} universe models. Specifically, we make no assumptions about the metric tensor of the space-time or the field equations prescribing it\footnote{In practice we must impose a minimal set of assumptions for the Taylor series expansion to be well defined on the space-time domain of interest. See \citep{Heinesen:2020b} for details.}. 
Here we state the final result of the detailed derivations, which can be found in \citep{Heinesen:2020b} along with a discussion of physical implications and application to the analysis of cosmological data. 
The luminosity distance in the vicinity of the observer is in this case given by the general expression %\hmt{, again up to third order in redshift, }
%\hmr{I removed the last term to be consistent with the FLRW equation} \ahr{I am a bit in favour of reinserting the last term, because of the special attention that error terms require in the inhomogeneous case.}\hmr{Ok, good point!}
\begin{equation}\label{eq:series}
    d_L(z) =  d_L^{(1)} z   + d_L^{(2)} z^2 +  d_L^{(3)} z^3 + \mathcal{O}( z^4),
\end{equation} 
with coefficients 
\begin{equation}
\begin{aligned}
\label{eq:dLexpand2}
d_L^{(1)} &= \frac{1}{\Eu_o} \, , \qquad d_L^{(2)} =   \frac{1 - \mathfrak{Q}_o }{2 \Eu_o}  \ ,\\% \qquad 
d_L^{(3)} &=  \frac{- 1 +  3 \mathfrak{Q}_o^2 + \mathfrak{Q}_o    -  \mathfrak{J}_o   + \mathfrak{R}_o }{ 6  \Eu_o}     \, , 
\end{aligned}
\end{equation} 
% The anisotropic parameters are given in terms of derivatives of the photon energy function, $E$, as measured in the frame of the observer, and the Ricci curvature tensor of space-time, $R_{\mu \nu}$, in the following way 
where the anisotropic parameters are defined as 
\begin{subequations}\label{eq:paramseff}
    \begin{align}
    %\hspace*{-0.7cm} 
    \label{eq:hdef}
    \Eu &= - \frac{1}{E^2}     \frac{ {\rm d} E }{{\rm d} \lambda}  \, ,\\ % \qquad 
    \mathfrak{Q}  &\equiv - 1 - \frac{1}{E} \frac{     \frac{ {\rm d} \Eu}{{\rm d} \lambda}    }{\Eu^2}   \, , \\ %\qquad
    \mathfrak{R} &\equiv  1 +  \mathfrak{Q}  - \frac{1}{2 E^2} \frac{k^{\mu}k^\nu R_{\mu \nu} }{\Eu^2}   \, , \\% \qquad
    \mathfrak{J}  &\equiv   \frac{1}{E^2} \frac{      \frac{  {\rm d^2} \Eu}{{\rm d} \lambda^2}    }{\Eu^3}  - 4  \mathfrak{Q}  - 3 \, .
    \end{align}
\end{subequations}
Here, $k^\mu$ is the 4-momentum of the incoming null ray, and the operator $\frac{ {\rm d}  }{{\rm d} \lambda} \equiv k^\mu \nabla_\mu$ is the directional derivative along the null ray with affine parameter $\lambda$. 
The photon energy function as measured by an observer with 4--velocity $u^\mu$ is $E\equiv-k^\mu u_\mu$, and $R_{\mu \nu}$ is the Ricci curvature tensor of {the} space-time. 
The set of anisotropic parameters $\{\Eu,\mathfrak{Q},\mathfrak{J},\mathfrak{R}\}$ formally enter the series expansion of $d_L$ in the same way as the FLRW parameters $\{H,q,j,\Omega_k \}$ in the expansion of $d_{L,{\rm FLRW}}$, and reduce to these parameters in the limit of exact homogeneity and isotropy. 
We thus denote $\{\Eu,\mathfrak{Q},\mathfrak{J},\mathfrak{R}\}$ the \emph{effective} observational Hubble, deceleration, jerk and curvature parameters. 
Comparing the parameters (\ref{eq:paramseff}) with their FLRW limits in (\ref{eq:dLFLRWdefs}), we see that $1/E$ plays the role of an effective \sayy{scale factor} on the null cone of the observer. 

The effective observational Hubble parameter $\Eu$ can be rewritten as a multipole series in the unit vector $e^\mu$ defining the incoming spatial direction of the null ray (i.e. the position of the astrophysical source on the sky as seen by the observer, see Appendix~\ref{sec:emu}):
\begin{equation}
\label{def:Eevolution}
\Eu(\boldsymbol{e}) = \frac{1}{3}\theta  - e^\mu a_\mu + e^\mu e^\nu \sigma_{\mu \nu}   \,  , 
\end{equation} 
where $\theta$ is the volume expansion rate, $\sigma_{\mu \nu}$ is the volume preserving deformation (shear tensor), and $a^\mu$ is the 4--acceleration of the observer congruence (see Appendix~\ref{appx:mesc} for mathematical definitions of these variables). 
{We note that the representation of $\Eu$ in (\ref{def:Eevolution}) is \emph{exact}, i.e., the truncation at quadrupolar order is a fundamental property of any observer congruence description and follows from the definition (\ref{eq:hdef}).} 
The multipole coefficients $\{\theta, - a_\mu, \sigma_{\mu \nu} \}$ represent 9 scalar degrees of freedom in total. 
The effective deceleration parameter can also be written {in \emph{exact} form} as an multipole series in $e^\mu$ as truncated at the order of the 16-pole: 
%\hmr{Decide if we want to keep this expression here. Maybe refer to it in the discussion around the sky-maps in Figure~\ref{fig:200_params_skymap}} \hmr{I changed the formatting of this eqn because it was overlapping the page and getting on my nerves - feel free to adjust as you like!} \ahr{Yes, I found that when writing both about the sky maps and about convergence, that it is good to have these equations to refer to!}
% \bea
% \label{q}
% \mathfrak{Q}(\boldsymbol{e} ) &=&  - 1 -  \frac{ \overset{0}{\mathfrak{q}}   +  e^\mu  \overset{1}{\mathfrak{q}}_\mu   +    e^\mu e^\nu   \overset{2}{\mathfrak{q}}_{\mu \nu}     +    e^\mu e^\nu e^\rho \overset{3}{\mathfrak{q}}_{\mu \nu \rho}    +   e^\mu e^\nu e^\rho e^\kappa  \overset{4}{\mathfrak{q}}_{\mu \nu \rho \kappa}   }{\Eu^2(\boldsymbol{e} )}    \, , 
% \eea 
\begin{equation}
\begin{aligned}
\label{q}
\mathfrak{Q}(\boldsymbol{e} ) = - 1& -  \frac{1}{\Eu^2(\boldsymbol{e} )} \bigg(
\overset{0}{\mathfrak{q}}   +  e^\mu  \overset{1}{\mathfrak{q}}_\mu   +    e^\mu e^\nu   \overset{2}{\mathfrak{q}}_{\mu \nu}     \\
&+    e^\mu e^\nu e^\rho \overset{3}{\mathfrak{q}}_{\mu \nu \rho}    +   e^\mu e^\nu e^\rho e^\kappa  \overset{4}{\mathfrak{q}}_{\mu \nu \rho \kappa} \bigg)     \, , 
\end{aligned}
\end{equation}
with coefficients 
\bea
\label{qpoles}
&& \overset{0}{\mathfrak{q}} \equiv  \frac{1}{3}   \frac{ {\rm d}  \theta}{{\rm d} \tau} + \frac{1}{3} D_{   \mu} a^{\mu  } - \frac{2}{3}a^{\mu} a_{\mu}    - \frac{2}{5} \sigma_{\mu \nu} \sigma^{\mu \nu}    \, , \nonumber \\ 
&& \overset{1}{\mathfrak{q}}_\mu \equiv  - \frac{1}{3} D_{\mu} \theta   -  \frac{2}{5}   D_{  \nu} \sigma^{\nu }_{\;  \mu  }   -  \frac{ {\rm d}  a_\mu }{{\rm d} \tau}  + a^\nu \omega_{\mu \nu}  +  \frac{9}{5}  a^\nu \sigma_{\mu \nu}     \, , \nonumber \\
&& \overset{2}{\mathfrak{q}}_{\mu \nu}  \equiv     \frac{ {\rm d}  \sigma_{\mu \nu}   }{{\rm d} \tau} +   D_{  \la \mu} a_{\nu \ra } + a_{\la \mu}a_{\nu \ra }     - 2 \sigma_{\alpha (  \mu} \omega^\alpha_{\; \nu )}   - \frac{6}{7} \sigma_{\alpha \la \mu} \sigma^\alpha_{\; \nu \ra }   \, , \nonumber \\ 
 && \overset{3}{\mathfrak{q}}_{\mu \nu \rho}  \equiv -  D_{ \la \mu} \sigma_{\nu   \rho \ra }    -  3  a_{ \la \mu} \sigma_{\nu \rho \ra }    \, , \nonumber\\% \qquad \quad  
 && \overset{4}{\mathfrak{q}}_{\mu \nu \rho \kappa}  \equiv  2   \sigma_{\la \mu \nu } \sigma_{\rho \kappa \ra} \, , 
\eea 
where the operator $\frac{ {\rm d} }{{\rm d} \tau} \equiv u^\mu \nabla_\mu$ is the directional derivative along the 4--velocity field $u^\mu$ of the observer congruence,   $\omega_{\mu \nu}$ is the vorticity tensor describing the rotational deformation of the observer congruence, and triangular brackets $\la \ra$ denote the traceless and symmetric part of the tensor in the involved indices. 
Similarly the effective observational curvature parameter $\mathfrak{R}(\boldsymbol{e})$ can be written as a series in $e^\mu$ truncated at the 16-pole level, whereas $\mathfrak{J}(\boldsymbol{e})$ is truncated at the order of the 64-pole -- see Appendix B in \citep{Heinesen:2020b} for the multipole series representations of $\mathfrak{R}(\boldsymbol{e})$ and $\mathfrak{J}(\boldsymbol{e})$.  
The coefficients of the multipole series expressions of $\{\Eu,\mathfrak{Q},\mathfrak{J},\mathfrak{R}\}$ are given in terms of physically interpretable kinematic variables and the curvature of space-time. % of the observer congruence
%As discussed in \citet{Heinesen:2020b}, 
Due to their truncated multipole series in $e^\mu$, the effective observational parameters are given in terms of a finite set of degrees of freedom. This property makes the described cosmographic representation of $d_L$ powerful for fully model independent analysis of cosmological data.  

%{We also note that 
The multipole coefficients that determine the effective observational parameters $\{\Eu,\mathfrak{Q},\mathfrak{J},\mathfrak{R}\}$ for each direction on the sky of the observer are covariantly given in terms of the kinematic variables $\theta$, $\sigma_{\mu \nu}$, $\omega_{\mu \nu}$ and the 4--acceleration $a^\mu$ of the observer congruence, and their covariant derivatives, along with space-time curvature invariants. % associated with the space-time. 
These covariant quantities are of fundamental interest in cosmology, and could be directly measured in,
%(without the need for model assumptions) 
e.g., large datasets of supernovae using this formalism. 
%data analysis} by for instance employing this formalism to the investigation of large datasets of supernovae. 

%\hmr{I'm a bit confused - a few paragraphs earlier you said that in the FLRW limit the parameters reduce to the FLRW parameters, and here in the following sentence you say there are non-trivial corrections? Perhaps I'm misinterpreting one (or both) of these statements...}
Inhomogeneities and anisotropies in general give rise to non-trivial corrections to the FLRW coefficients of \eqref{eq:series}: the effective observational parameters $\{\Eu,\mathfrak{Q},\mathfrak{J},\mathfrak{R}\}$ are sensitive to the local environment of the observer and the direction of the astrophysical source. 
Even though the parameters $\{\Eu,\mathfrak{Q},\mathfrak{J},\mathfrak{R}\}$ formally replace the FLRW parameters $\{H,q,j,\Omega_k \}$ in the series expansion of luminosity distance, their physical interpretation are in general \emph{not} those of the FLRW space-time. For instance, $\mathfrak{Q}$ does not in general measure the physical deceleration of space, and is thus not necessarily positive in a decelerating universe model. Similarly, $\mathfrak{R}$ does not in general relate simply to the Ricci curvature of spatial sections, as $\Omega_k$ does in FLRW. 
%measure the spatial curvature over canonical cosmological volume sections and is in general not restricted in its sign by the sign of the spatial Ricci scalar as defined in the observer frame. }

For the series expansion \eqref{eq:series} to be well defined, we require the effective Hubble parameter $\Eu$ to be differentiable and of constant sign everywhere in the domain of application of the series \citep[see][for a detailed discussion on regularity requirements of the general series expansion]{Heinesen:2020b}.
{For expanding universe models this translates into the requirement of positivity of $\Eu$,
%{(for realistic universe models, where most of the volume in the frame of the matter is expanding, this translates into the requirement of positive sign of $\Eu$)}
%The requirement of positive sign of $\Eu$ 
which itself} effectively implies the need to impose a coarse-graining scale above that of
% the scale of 
the largest collapsing regions\footnote{{In 
%expanding \aht{(or contracting)} 
FLRW {universe} models, the requirement of constant sign of $H$ is satisfied per construction. %In FLRW space-times, problems related to coarse-graining do not arise as these have already been dealt with in an implicit manner by simply hypothesising a space-time metric ansatz valid on the largest scales. 
However, in the presence of inhomogeneity and anisotropy, care must be taken to ensure the requirement of a well behaved series expansion \eqref{eq:series}.}}. 
%{The requirement $\Eu \neq 0$ over the space-time domain of interest 
This ensures that the redshift function is monotonic along the null ray, 
%s. 
%This monotonicity requirement must be satisfied in order for 
such that there exists a cosmological notion of a distance-redshift relation, i.e., for distance to be a single valued function of redshift. 
In the context of our work, 
%\aht{the imposing of a coarse-graining scale}
imposing a coarse-graining scale 
translates to employing a smoothing procedure to exclude small-scale nonlinear dynamics in our simulations, which we explain further in the next section.

%We also wish to stress that this requirement is not unique to the generalised cosmography, and also applies to the FLRW cosmography summarised in Section~\ref{sec:FLRWdL}. However, the FLRW assumption (and its typical application to expanding space-times) explicitly ensures positivity of $H_0$. In the presence of inhomogeneity and anisotropy, care must be taken to ensure this requirement is met.

%\hmr{Is this true? Feel free to re-word however you like.Maybe we can say something w.r.t surveys, i.e. note that so long as the series is convergent in a given z range this requirement doesn't affect its applicability to data} 
%\ahr{I now inserted a footnote with these comments above and also some comments of what the requirement physically implies for the model universe. }

%In the next section, we explain how we meet these requirements in our simulations.
%\footnote{For a detailed discussion on regularity requirements involved in the formulation of the series expansion \eqref{eq:series} and their physical implications, see \citet{Heinesen:2020b}. In the next section, we explain how we meet these requirements in our simulations.}

\section{Numerical relativity simulations}\label{sec:sims} 

In this work, we wish to remain agnostic about the existence of a global background cosmology of any kind, or the smallness of any perturbations. 
We therefore use numerical relativity simulations, which contain no {a priori imposed} physical constraints on the form of the metric tensor.
This allows us to mimic as closely as possible the generality of the formalism presented in Section~\ref{sec:GeneraldL}. 

We now present the general relativistic numerical simulations used in our analysis. In Section~\ref{sec:software} we describe the software used and the physical assumptions about the energy momentum tensor. In Section~\ref{sec:ics} we describe the initial conditions of the simulations, chosen to be consistent with the $\Lambda$CDM matter power spectrum at early times following the recombination epoch. % \ahr{Correct?} 
In Section~\ref{sec:postprocessing} we describe volume average properties of the simulated space-time, and the observers analysed in the subsequent analysis.

%\ahr{I added another subsection together with a small introduction to make the structure between the main sections more uniform.}\hmr{great!}

\subsection{Software and physical assumptions} \label{sec:software}

%\ahr{The following sentence is long (some things seems to not make sense gramatically?) Break.}\hmr{thanks - I removed some unnecessary stuff}
We use the Einstein Toolkit\footnote{\url{https://einsteintoolkit.org}} \citep[ET;][]{Loffler:2012,Zilhao:2013}, a free, open-source numerical relativity code based on the Cactus\footnote{\url{https://cactuscode.org}} infrastructure. The ET comprises modules (``thorns'') to evolve the Einstein equations using the BSSNOK formalism \citep{Nakamura:1987,Baumgarte:1999,Shibata:1995} alongside the equations of general-relativistic hydrodynamics, while also providing thorns for initial data,  analysis, and the handling of adaptive mesh refinement and MPI. The ET has long been used for, e.g., simulations of compact relativistic objects and the emission of gravitational waves. Recently, the ET has also been applied to simulations of inhomogeneous cosmology \citep{Bentivegna:2016stg,Macpherson:2016ict,Macpherson:2019a,Wang:2018a}, and has proven to be a reliable tool to study nonlinear structure formation.

In this work, we use the \texttt{McLachlan} thorn \texttt{ML\_BSSN} \citep{Brown:2009} in combination with \texttt{GRHydro} \citep{Baiotti:2005,Mosta:2014} to evolve cosmological initial data provided by \texttt{FLRWSolver} \citep[][see also Section~\ref{sec:ics} below]{Macpherson:2016ict}. We refer the reader to \citep{Macpherson:2018akp} for further specifics on the simulations. However, we note here that \texttt{GRHydro} adopts a fluid approximation (i.e., no collisionless particles), and that our simulations contain no dark energy ($\Lambda=0$). 
%\hmr{I think the following paragraph is not necessary, I will just point to my paper where we discuss this anyway}\texttt{GRHydro} adopts a fluid approximation (i.e., no collisionless particles) with equation of state (EOS) handled by the thorn \texttt{EOS\_Omni}. Currently, there is no option to simulate pure dust, i.e. $P=0$. Instead, we choose an EOS such that the pressure $P\ll\rho$, which was shown to be accurate in \citet{Macpherson:2016ict} for simulating a dust FLRW cosmology. We do expect the small amount of pressure in our simulations to have an effect on small scales when structures are nonlinear, by way of slowing the collapse of these structures compared to a pure-dust evolution.
%Our simulations use periodic boundary conditions and contain no dark energy ($\Lambda=0$), the latter purely because the evolution thorns we use were not written for cosmological applications. However, see \citet{Bentivegna:2016stg} for a collection of cosmology-specific thorns, \texttt{CTThorns}, to evolve a pure-dust cosmology with a cosmological constant. We intend to integrate \flrwsolver\, with this collection of thorns in the future, however in this work we are interested in the ability of a matter-dominated universe to mimic the behaviour of one dominated by a cosmological constant. 
%\ahr{Maybe a sentence here about the convergence of the simulation on the largest scales to the EdS background?}\hmr{done - I do mention this in the post-processing section below, but added a sentence here too, after the next sentence}
We therefore compare our numerical calculations to the flat, matter-dominated Einstein-de Sitter (EdS) model (in line with our initial conditions, see Section~\ref{sec:ics} below). While not explicitly enforced, we find that our simulations converge to EdS behaviour when averaged over the largest scales \citep[see Section~\ref{sec:postprocessing} below and also][for a set of similar simulations]{Macpherson:2019a}.

%\ahr{I have put the below paragraph in color, since I am now thinking of removing it. We can say that we are applying a conservative coarsegraining scale at and above that of FLRW statistical homogeneity scale, avoiding this technical discussion?.} \hmr{I moved part of this paragraph to the previous section, and removed the technical discussion here.}

%\hmt{To ensure the validity of applying the series expansion \eqref{eq:series}, we must ensure that our simulations meet the same criteria outlined at the end of the previous section.}. To this end, we apply a conservative coarsegraining scale at and above that of the FLRW statistical homogeneity scale.
%Calculating averages of tensorial quantities (e.g., the shear tensor in this case) is a non-trivial and non-unique operation. To avoid such complications we perform simulations containing \emph{only} large-scale structure, such that no collapse (negative expansion) occurs anywhere in the domain, and such that shear and 4-acceleration of the observer congruence are subdominant relative to the volume expansion rate. This ensures that the necessary condition $\mathfrak{H} > 0$ for the series expansion \eqref{eq:series} is satisfied in the simulation. To this end, each individual grid cell in our simulations coincides with the smoothing scale of interest.
Our simulations have resolution $N^3$ with $N=128$, and we choose physical coarse-graining scales such that individual grid cells have length $100 \,h^{-1}{\rm Mpc}$ and $200 \,h^{-1}{\rm Mpc}$, corresponding to total domain lengths of $12.8\, h^{-1}{\rm Gpc}$ and $25.6\, h^{-1}{\rm Gpc}$, respectively. We also perform lower-resolution simulations with $N=32$ and $64$ to show that our results are robust to changes in resolution (see Appendix~\ref{appx:errors}). 
We choose this method of coarse graining to \emph{strictly} exclude small-scale nonlinearities, 
%while also avoiding issues associated with smoothing tensorial quantities \hmr{add a cite}. 
to ensure the positivity requirements of $\Eu$ in the generalised series expansion are met {(see discussion in Section~\ref{sec:GeneraldL})}.
However, in Appendix~\ref{appx:errors} we present simulations with a resolution-independent smoothing procedure and note that we find consistent results. 
%In addition, in Appendix~\ref{appx:series_convergence} we discuss the convergence and quality of the approximation of the general series expansion in the context of these simulations. %I moved this to an earlier section
%, in order to justify the range of redshifts over which our results are applicable.

Our chosen coarse-graining scales are, respectively, comparable to and larger than the estimated \sayy{1\%} statistical homogeneity scale of $\sim 100 h^{-1}$ Mpc measured from galaxy catalogues \citep{Hogg:2004vw,2012MNRAS.425..116S}. At these scales, typical density contrasts are expected to be $\sim 1\%$ at the present epoch in the $\Lambda$CDM model.
Note however that our simulations will have larger density contrasts both 
%Note however that the \sayy{present epoch} density contrasts are of order $\sim 10\%$ at smoothing scales $(100 \,{\rm Mpc})^3$ and $(200 \,{\rm Mpc})^3$ in the present simulations, 
due to the absence of a cosmological constant and to our volume-based definition of the \sayy{present epoch} in our simulations (see Section \ref{sec:postprocessing}). 
The coarse-graining scales of our analysis can nevertheless be considered conservative, as density contrasts are still within the linear regime in which the FLRW metric is usually considered safe as a 
%to employ for a good 
lowest order description. 

\begin{figure*}%[h]
    %\centering
    \includegraphics[width=0.8\textwidth]{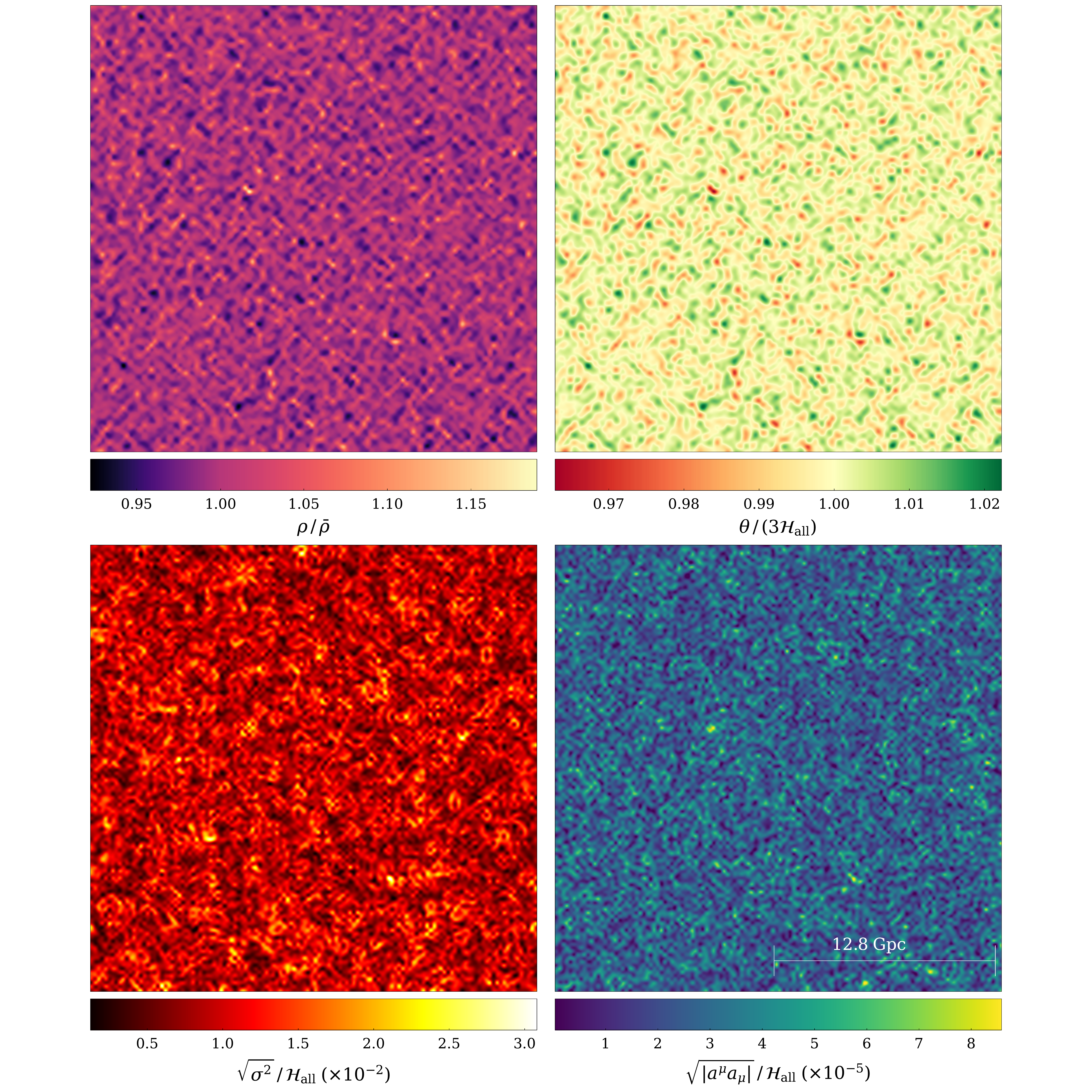}
    %{128_25p6Gpc_2x2_2Dslice.pdf}
    \caption{Rest-mass density, expansion rate, shear, and acceleration (panels; top left to bottom right, respectively) in a $128^3$ resolution simulation with a box size of $25.6 h^{-1}$Gpc. Each panel shows a 2--dimensional slice through the 3--dimensional domain at $\zeff=0$. 
    }
    \label{fig:2Dslices}
\end{figure*}

Figure~\ref{fig:2Dslices} shows the 25.6 $h^{-1}$ Gpc, $N=128$ simulation used in this work. Panels, top left to bottom right, show the rest-mass density field relative to the global average, the expansion scalar, shear scalar $\sigma^2\equiv \frac{1}{2} \sigma^{\mu\nu}\sigma_{\mu\nu}$, and 4--acceleration magnitude each normalised by the global expansion, $\mathcal{H}_{\rm all}$.
We show 2--dimensional slices through the midplane of the domain at effective redshift $\zeff=0$. See Section~\ref{sec:postprocessing} below for definitions of $\zeff$ and $\mathcal{H}_{\rm all}$. 
%The length of the domain here is 25.6 $h^{-1}$ Gpc, such that individual grid cells have length 200 $h^{-1}$Mpc.  
%\ahr{Right now we have two different normalisations $H_{all}$ and $H_{EdS}$ which are practically the same. Perhaps use $H_{EdS}$ everywhere?}\hmr{I agree - but I would rather keep Hall in the simulations plot, since then it is just a plot of the simulation and not w.r.t some reference model. I'll try to make it more consistent elsewhere though!}
The expansion scalar, shear tensor, and 4--acceleration shown here are the quantities subsequently used to calculate the observational effective cosmological parameters.

%This statistical homogeneity scale is often used in practice as a scale beyond which the FLRW metric model applies to our observed data.

\subsection{Initial data and numerical gauge}\label{sec:ics} 

%\ahr{In this section, put definition of fluid/observer 4-velocity}
%\hmr{I'll put this in the next section, since it is related to the observers which are placed in post-processing}

\flrwsolver\, is a thorn for the ET\footnote{\flrwsolver\, is not yet available as a part of the ET, however you can find the public version at \url{https://github.com/hayleyjm/FLRWSolver_public}.} developed by \citet{Macpherson:2016ict} to generate and implement initial conditions for the linearly-perturbed, flat FLRW metric in longitudinal form, namely 
\begin{equation}\label{eq:longmetric}
    ds^2 = - a(\eta)^2 (1 + 2\phi) d\eta^2 + a(\eta)^2 (1 - 2\phi) \delta_{ij} dx^i dx^j,
\end{equation}
where $\eta$ is conformal time, $a(\eta)$ is the background scale factor, and $\phi\ll1$.
%\flrwsolver\ uses a given matter power spectrum to generate 
The initial density fluctuations are Gaussian-random and drawn from a user-provided matter power spectrum. 
We use CLASS\footnote{http://class-code.net} to generate the input matter power spectrum (in the longitudinal gauge) at redshift $z=1000$, with $h=0.7$ and otherwise default parameters \citep{CLASS:2011}. 
The corresponding metric perturbation $\phi$ in \eqref{eq:longmetric} and the velocity field $v^i$ are calculated via the linearised Einstein equations in the EdS model \citep[see][for details]{Macpherson:2016ict}. Sampled scales depend on the physical size of the domain and the numerical resolution: the largest mode sampled is the physical side length of the total cubic domain, and the smallest mode is $2\times$ the physical size of the grid cells\footnote{We also perform simulations with structures below $10\times$ the grid cell size cut out (see Appendix~\ref{appx:pkcutsims}), and find similar results.}. 
{The length scales quoted in our paper in units of Mpc/h are given in terms of our choice of $h=0.7$ in CLASS. However, $h=0.7$ does not correspond to the present epoch Hubble parameter of our simulations, which is on average well approximated by the Hubble parameter of an EdS space-time (see Section~\ref{sec:postprocessing}).}

The metric is assumed to be of the form \eqref{eq:longmetric} on the initial hypersurface, and implicitly for the first few time steps in order to specify
%because we must also specify 
the extrinsic curvature $K_{ij} \propto \pd_t \gam_{ij}$ on the initial slice. 
%, which we take to be linear in $\phi$.
Throughout the simulation, however, the metric is best described in the general $3+1$ form
\begin{equation}\label{eq:ds}
    ds^2 = -\alpha^2 dt^2 + \gamma_{ij} dx^i dx^j,
\end{equation}
where $t$ is coordinate time, $\alpha$ is the lapse function
%\ahr{I would suggest replacing the content of the following parenthesis with: 'representing the freedom of choice of coordinate time t'} 
%(describing the spacing between subsequent \hmt{constant-time} slices)
(representing the freedom of choice of coordinate time $t$), $\gamma_{ij}$ is the spatial metric, and we have forced the shift vector (describing the shift in spatial coordinates between subsequent time slices) $\beta^i=0$ throughout the simulation for convenience.  %as a convenient gauge choice. 
We choose a harmonic-type 
%gauge\footnote{We note the term \sayy{gauge} here is used in the numerical relativity sense --- i.e., that of coordinate evolution ---, not in the typical sense for which it is used in cosmology --- i.e., that of defining a particular form of the metric.} for the 
evolution of the lapse function, namely
%\begin{equation}\label{eq:gauge}
%\partial_t \alpha = - \frac{1}{3}\alpha^2 K
$\partial_t \alpha = - \alpha^2 K/3$,
%\end{equation}
where $K=K^i_{\phantom{i}i}$ is the trace of the extrinsic curvature. In the linear regime, this translates to $\alp'/\alp = a'/a$, i.e. $\phi'=0$ for our choice of initial metric \eqref{eq:longmetric}, and therefore equates to choosing the pure growing mode of the linear density perturbation, $\delta$ \citep[see][]{Macpherson:2016ict}. Once the perturbations grow nonlinear, this interpretation of our gauge choice is no longer applicable, nor is the metric \eqref{eq:longmetric}.

\subsection{Post-processing analysis}\label{sec:postprocessing}

We use a new version of the analysis code \mesc\ \citep[written specifically for ET data, see][]{Macpherson:2019a} to calculate the terms in the series expansion \eqref{eq:dLexpand2}.
\texttt{Mescaline} adopts a uniform Cartesian grid with periodic boundary conditions, i.e. employing a torus condition on the topology of the spatial sections. 
Otherwise, the code is completely general, with no physical assumptions on the form of the metric or the fluid model of the energy momentum tensor. 
%\hmr{actually, the code might assume dust in some parts. not for our calculations, but for e.g. the backreaction calculation it might. so I will clarify this sentence if so...}\hmr{just checked, and it doesn't explicitly in the form of the eqns. in the sense that it doesn't read in any pressure data, I guess it does assume P=0, but this is nowhere explicitly enforced, i.e. P wouldn't arise anywhere in the calculations anyway.}

%\hmr{note I've shuffled this section about a bit to put the relevant calculations details first}
We place 1000 observers at pseudo-randomly chosen positions within the simulation domain. The observers are co-moving with the fluid flow, such that %the observers 
they are moving along world lines defined by the fluid 4--velocity $u^\mu \equiv \frac{dx^\mu}{d\tau}$,
where $\tau$ is the proper time. In terms of $3+1$ variables, the 4--velocity can be split into its time and space components
$u^0 = \Gam/\alp$, and $u^i = \Gam v^i$, respectively (for $\beta^i=0$). 
% \begin{equation}
%     u^0 = \frac{\Gam}{\alp}, \quad u^i = \Gam v^i,
% \end{equation}
Here, $v^i$ is the 3--velocity with respect to the Eulerian observer, and $\Gam=1/\sqrt{1-v^iv_i}$ is the Lorentz factor. We note the distinction between the fluid 4--velocity $u^\mu$ --- the physical time direction
%\hmr{isn't this also a physical spatial flow?} \ahr{A 4-velocity field is per construction defining the time-arrow of the observer in question. Now of course, if you see it from another frame, there will in general be a spatial component, just because of how SR frames are related}
of the fluid flow in the simulation --- and the normal vector $n^\mu$ describing the arbitrary foliation of the simulation. 
%\ahr{I think that your above explanation is sufficient, and that the following sentence can be left out?}
%\hmt{Since our simulations are not performed in the frame co-moving with the fluid flow, we choose physically motivated observers who are tilted with respect to our chosen foliation.}  

\begin{figure}%[h]
     \centering
     \begin{subfigure}[b]{0.9\columnwidth}
         \centering
         \includegraphics[width=0.9\columnwidth]{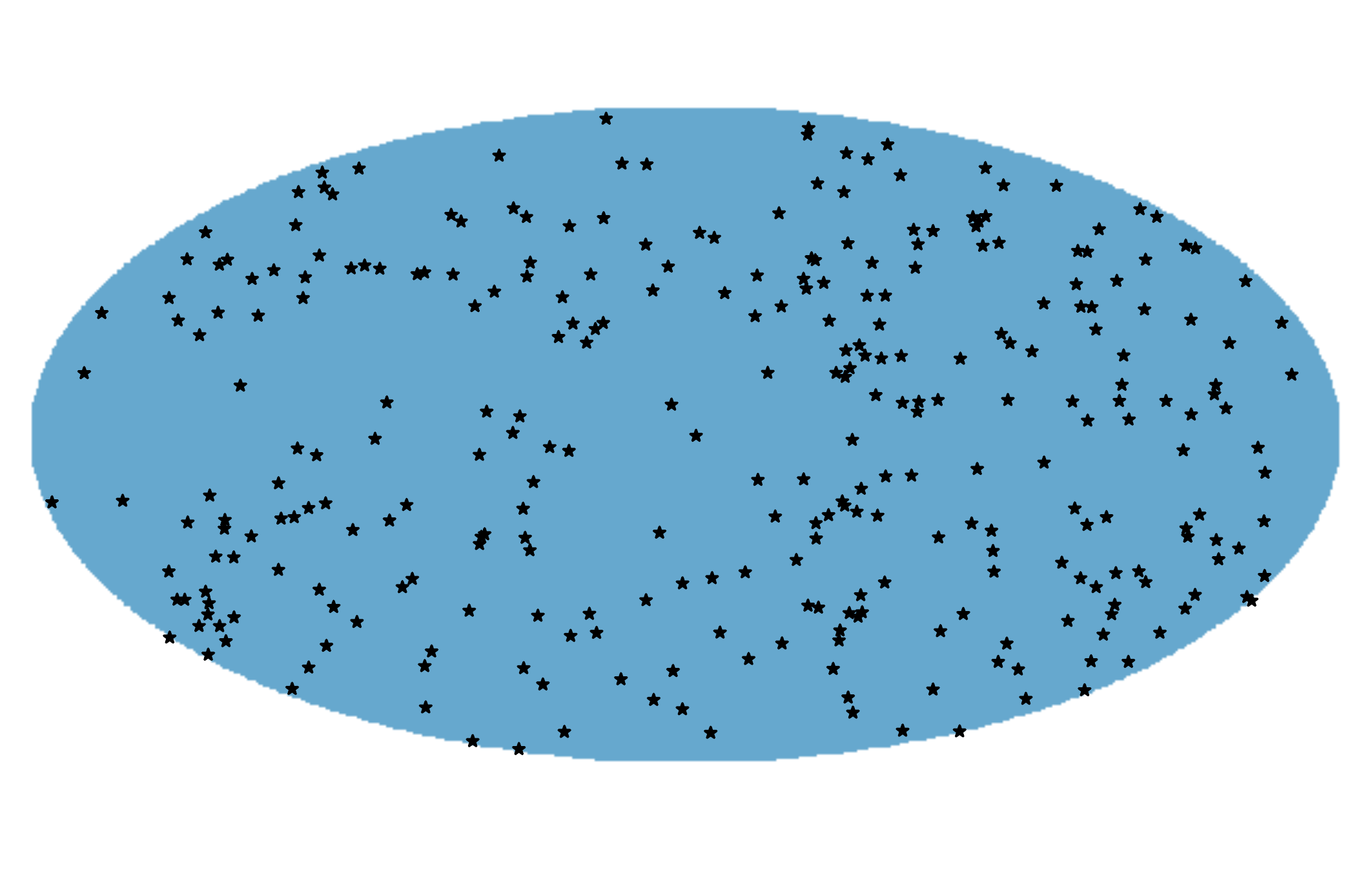}
         \caption{`FullSky' sample.}
         \label{fig:fullsky}
     \end{subfigure}
     \hfill
     \begin{subfigure}[b]{0.9\columnwidth}
         \centering
         \includegraphics[width=0.9\columnwidth]{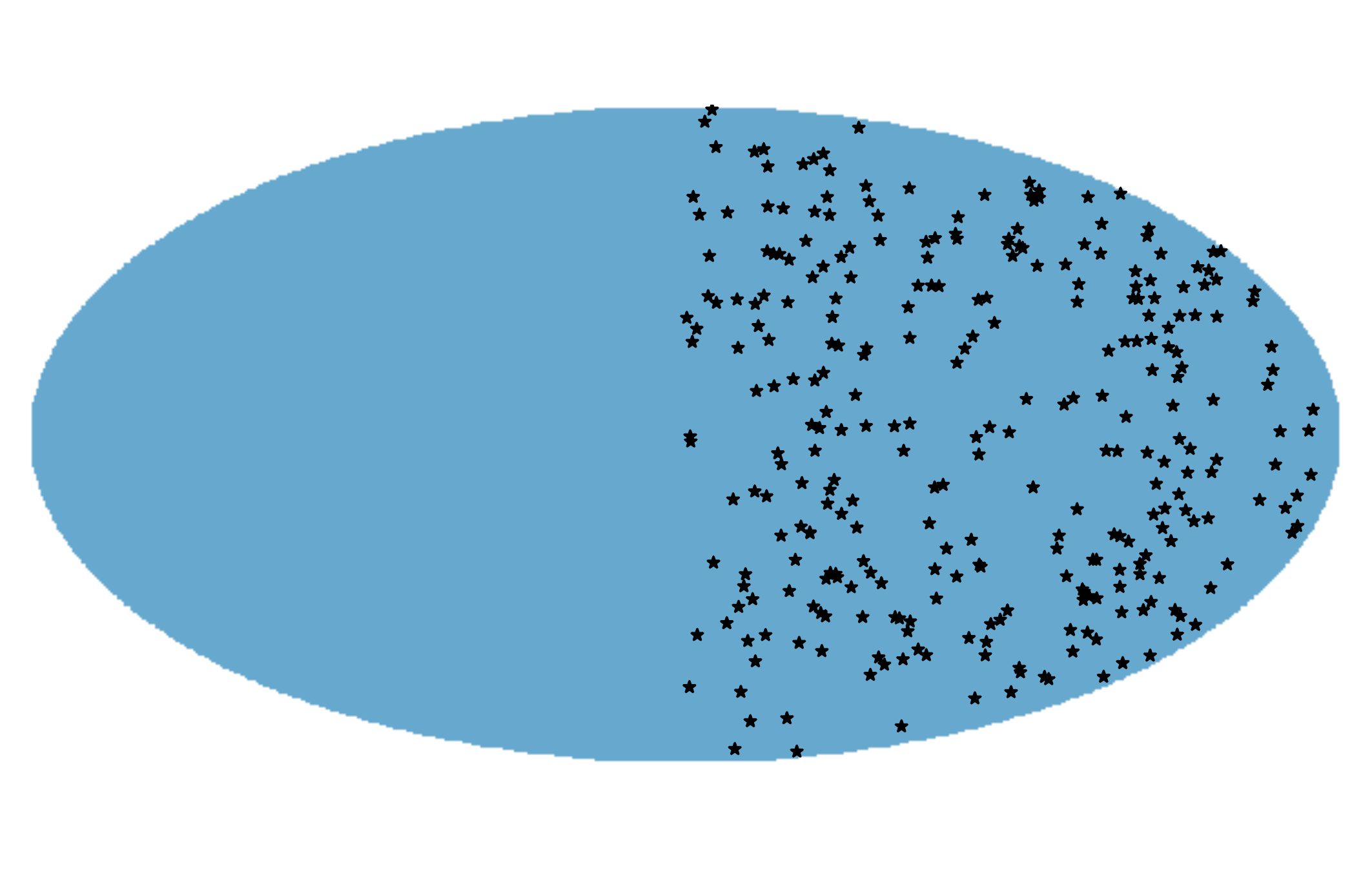}
         \caption{`HalfSky' sample.}
         \label{fig:halfsky}
     \end{subfigure}

        \caption{Example sky-maps for one observer with 300 lines of sight, representing the approximately isotropic (top) and anisotropic (bottom) sky coverage used in this work.}
        \label{fig:skymaps}
\end{figure}

For each observer, we choose 300 pseudo-random lines of sight \citep[similar to the number of SNe used in the local measurement of $H_0$ in][]{Riess:2016}. 
We focus on two different examples of sky-sampling for each observer -- see Appendix~\ref{sec:emu} for details on how the sky-samplings are generated.  
First is the case of a \sayy{fairly-sampled} sky, in which the lines of sight are chosen pseudo-randomly across the observer's whole sky. Figure~\ref{fig:fullsky} shows an example of this distribution for one observer, which we refer to in the text and figures as the \sayy{FullSky} sample. Second, we consider an \sayy{unfairly-sampled} sky, in which the lines of sight are chosen pseudo-randomly across \emph{one half} of the observer's sky. Figure~\ref{fig:halfsky} shows an example of this distribution, which we refer to in the text and figures as the `HalfSky' sample. 
%\ahr{By eye the Half sky seems more sampled in the upper part, but this could well be a random effect of course..}\hmr{I don't quite understand this yet -- so I will try to, and maybe add something}

For each line of sight, $e^\mu$, of the sky-maps  %(unique $e^\mu$, see Section~\ref{sec:emu}) 
we calculate the observational effective Hubble parameter \eqref{def:Eevolution} using $\theta$, $\sigma_{\mu\nu}$ and $a^\mu$ as evaluated at the observer's position. The remaining effective cosmological parameters \eqref{eq:paramseff} are subsequently built using the Ricci tensor and derivatives of $\Eu$. See Appendix~\ref{appx:mesc} for details of these calculations, including a consistency test using an analytic metric. 

In this work we focus on calculations of the effective cosmological parameters, rather than $d_L(z)$ itself. 
However, in Appendix~\ref{appx:series_convergence} we discuss the quality and convergence of the approximation of the general series expansion \eqref{eq:series} in the context of the simulations used here. 

%In addition to the terms in the general series expansion, 
We also use \mesc\ to assess the average dynamics of the simulation relative to the EdS model. This includes a calculation of the effective scale factor,
\begin{equation}\label{eq:aD}
    a_\mathcal{D} (t) \equiv \left( \frac{V_\mathcal{D} (t)}{V_{\mathcal{D},{\rm ini}}} \right)^{1/3},
\end{equation}
where $V_\mathcal{D}(t) \equiv \int_\mathcal{D} \sqrt{\gam} d^3 X$ is the volume of a domain $\mathcal{D}$ on the spatial surfaces, $\gam$ is the determinant of the spatial metric describing these surfaces, and $V_{\mathcal{D}, {\rm ini}}\equiv V_\mathcal{D}(t_{\rm ini})$ is the volume on the initial slice, with $a_{\mathcal{D},{\rm ini}} =1$ for a given domain \cite[for full details on the averaging procedure, see][]{Macpherson:2019a}. 
We also define the ``effective redshift'', $z_{\rm eff}$, in order to find appropriate spatial surfaces in which to place our observers\footnote{See \citet{Rasanen:2008be,Rasanen:2009uw} for plausible arguments for $\zeff$ as a good lowest order approximation for the measured redshift in space-times with slowly evolving structure and \emph{statistical} homogeneity and isotropy.}. Specifically, this is defined from \eqref{eq:aD} with $\mathcal{D}$ taken to be the total simulation domain, i.e. 
%\begin{equation}
%      a_{\mathcal{D}, {\rm all}} = \frac{z_{\rm ini} + 1 }{ \zeff + 1 },
%\end{equation} 
\begin{equation} 
    \zeff(t) + 1 \equiv \frac{a_{\mathcal{D}, {\rm all}}(t_{\rm 0}) }{a_{\mathcal{D}, {\rm all}}(t) }  , 
\end{equation}
where $a_{\mathcal{D}, {\rm all}}(t_{\rm 0}) \equiv 1000$ is the value of the scale factor defining the \sayy{present epoch} surface with $t=t_{\rm 0}$, which arises from our choice of $\zeff(t_{\rm ini}) + 1 = 1000$.
%where $z_{\rm ini}=1000$ is the initial redshift of the simulations, when $a_{\mathcal{D}, {\rm all}}=1$. 
%We use $\zeff$ to define surfaces in which to place observers
%, and as an indicative value of the mean redshift function associated with emitters and observers placed on the constant $t$ surfaces
%, and note that it is indicative \emph{only} of how much the volume of the simulation has increased, and is not an observational redshift 
%\hmr{I've highlighted the previous part of the sentence because, while I think it is an interesting point, I'm not sure it is relevant - because we really do only use $z_{\rm eff}$ to place observers, and not so much the latter part, do you agree?} 
%\ahr{Yes. I have now but the references to Syksy in a footnote, since I think it is still relevant for the physical intuition of the surfaces that this is indicative of the redshift function.}
We also define the global average expansion rate,
\begin{align}
    \mathcal{H}_{\rm all} &\equiv \frac{1}{3} \langle \theta \rangle_{\mathcal{D}, {\rm all}} \label{eq:huball} \equiv \frac{1}{3} \frac{1}{V_{\mathcal{D}, {\rm all}}} \int_{\mathcal{D}, {\rm all}} \theta\, \sqrt{\gam}\, d^3X,
\end{align}
where $\langle \rangle_{\mathcal{D}, {\rm all}}$ is the average over the entire simulation domain. %$\mathcal{D}$ --- here taken to be the entire simulation \hmt{domain}. % ---, $\langle \theta \rangle_\mathcal{D} = \frac{1}{V_\mathcal{D}} \int_\mathcal{D} \theta\sqrt{\gam} d^3X$. 
In our $N=128$ simulations, the globally-averaged expansion rate at the present epoch, $\mathcal{H}_{\rm all}(\zeff=0)$,
%\footnote{We use $\zeff\approx0$ to denote the simulation snapshots used because the simulation output lies close to but not \emph{exactly} coinciding with $a_{\mathcal{D}, {\rm all}}=z_{\rm ini}+1$.}
coincides with the EdS value, $\mathcal{H}_{0,{\rm EdS}}$, to within $1\%$. % note this is still true for new sims

Cosmological parameters averaged over the whole domain for this simulation are %$\Omega_m \approx 1.02$, $\Omega_R \approx -2 \times 10^{-4}$, and $\Omega_Q \approx 1.7 \times 10^{-6}$ <-- old sim
{$\Omega_m \approx 1.02$, $\Omega_R \approx 1 \times 10^{-5}$, and $\Omega_Q \approx -6 \times 10^{-9}$} %<-- new sim
for matter, curvature, and backreaction, respectively \citep[see][for details on the calculation of these parameters]{Macpherson:2019a}. %., coinciding with the EdS model. 
Therefore, in terms of the cosmological parameters, the simulations converge to the EdS background model at the scale of the entire simulation domain with an accuracy of $\sim 2\%$. This deviation from the background EdS model value of $\Omega_m = 1$ can be assigned to numerical errors (see Appendix~\ref{appx:errors}). 

%\ahr{These parameters don't sum to 1 at the level of accouracy of the simulations?!. Do you mean $\Omega_m \approx 1.0002$ ? And did you mean to cite Macpherson:2019a for the review of cosmological parameters?}\hmr{No, this value is correct! The remainder here is essentially the violation in the Hamiltonian constraint, plus any additional error from the calculation. The (dimensionless) Hamiltonian violation is a few \%, from a separate calculation, so this matches up! And yes, I have changed the reference - thanks!} %Our results therefore represent the effect of local inhomogeneities and anisotropies in a model universe that coincides with EdS on average.

%=============================================
\section{Results and discussion}\label{sec:results}

Here we present the main findings of our analysis. In Section~\ref{sec:skyvar} we analyse anisotropies of effective cosmological parameters of the luminosity distance across the individual observer's skies. In Section~\ref{sec:skyav} we consider the same cosmological parameters  averaged over the observer's skies for two simple choices of survey geometries and analyse the variance between observers. In Section~\ref{sec:H} we discuss the implications of our analysis for the Hubble tension present in the $\Lambda$CDM paradigm.

\subsection{Sky variance of observational effective parameters}\label{sec:skyvar}

%\subsubsection{Sky distributions}

\begin{figure*}%[h]
    \centering
    \includegraphics[width=\textwidth]{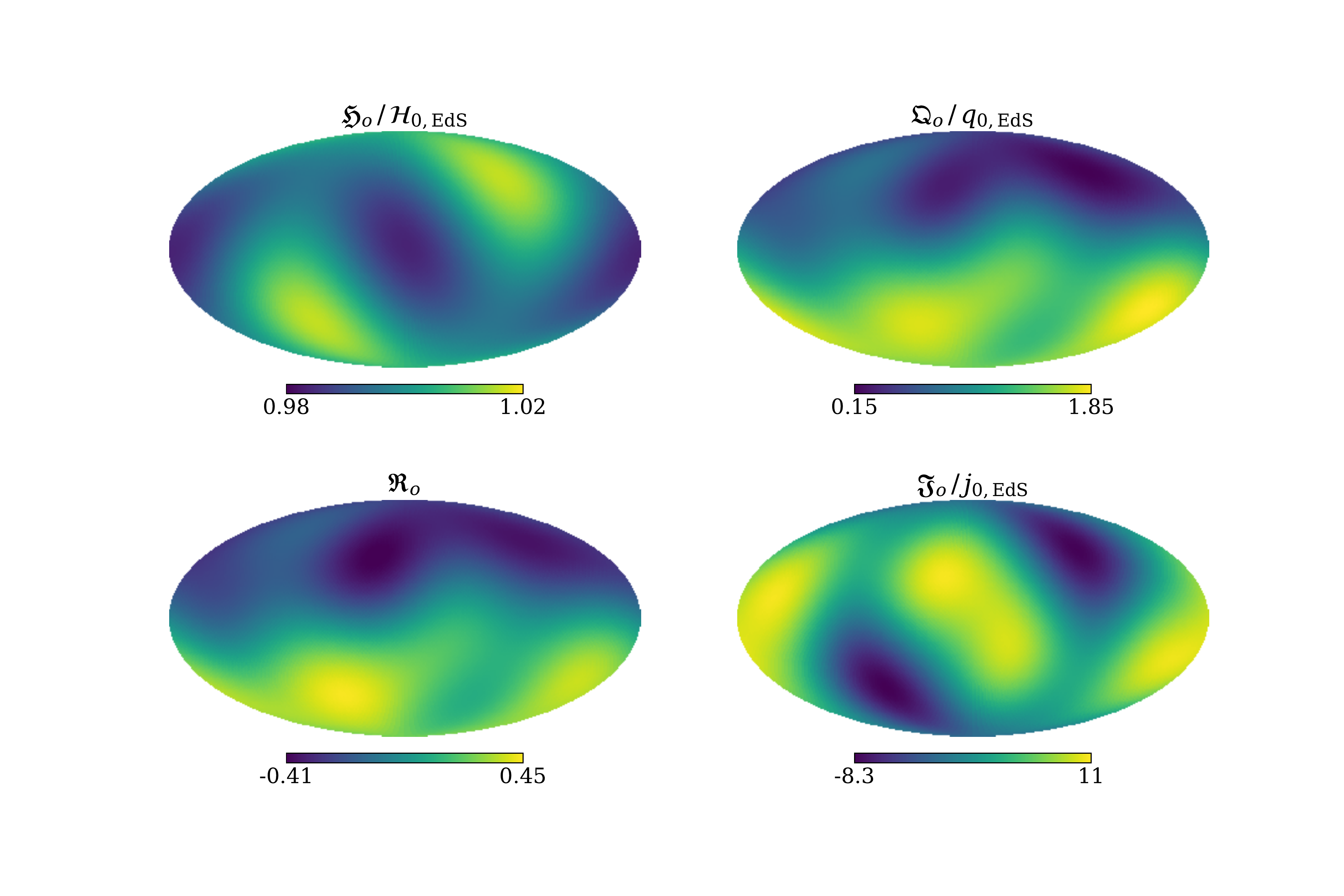}
    %{200Mpc_params_skymap_128_Healpix_obs6.pdf}
    \caption{Sky-maps of the effective Hubble, deceleration, curvature, and jerk parameters (top-left to bottom-right, respectively) for one observer measured in directions of the $12\times N_{\rm side}^2$ \texttt{HEALPix} pixels with $N_{\rm side}=32$. Each parameter (with the exception of the curvature) is normalised by its respective EdS value.
    }
    \label{fig:200_params_skymap}
\end{figure*}

Figure~\ref{fig:200_params_skymap} shows sky-maps of the observational effective Hubble, deceleration, curvature, and jerk parameters \eqref{eq:paramseff}, relative to their respective EdS values (top left to bottom right, respectively). We show maps measured by a single observer with lines of sight\footnote{We note that for this observer we have used a larger number of lines of sight than our main results in order to create smoother maps. All results are presented with 300 lines of sight.} in directions of the $12\times N_{\rm side}^2$ \texttt{HEALPix}\footnote{http://healpix.sf.net} \citep{Gorski:2005} pixels for $N_{\rm side}=32$. We note the distinction between the resolution of the sky-maps, $N_{\rm side}=32$, and the resolution of the simulation in which we place the observers, $N=128$.  %\hmr{maybe this could be a footnote, but we already have a lot on this page...}

%\ahr{potentially convert the following remark of number of sampled directions in terms of angular resolution or leave out?} with 9,000 isotropic lines of sight \hmr{todo: Sort out healpix directions properly} (maps were generated using \texttt{HEALPix}\footnote{http://healpix.sf.net}).
For this particular observer, we notice that the dominant form of anisotropy in the effective Hubble parameter is the quadrupole, i.e., the contribution from the shear tensor dominates the 4--acceleration term in \eqref{def:Eevolution}. 
The dipole is the dominating anisotropy in the effective deceleration parameter, which can be attributed to the two first terms of $\overset{1}{\mathfrak{q}}_\mu$ in \eqref{qpoles} involving the spatial gradient of the expansion rate and of the shear tensor, respectively.
The octopole moment is also visible in the angular distribution of the effective deceleration parameter, which can be assigned to the first term of $\overset{3}{\mathfrak{q}}_{\mu \nu \rho}$ involving the spatial gradient of the shear tensor. 
%\hmr{I brought the sentence in colour back, because it now seems more relevant for this observer?} \ahr{Yep I agree!}
The effective curvature parameter has similar angular distribution to the effective deceleration parameter, which is due to $\mathfrak{Q}$ entering the definition of $\mathfrak{R}$ in \eqref{eq:paramseff} where it dominates the anisotropic signal. 
The quadrupole dominates the effective jerk parameter, which can be assigned to the second spatial gradient of the expansion rate entering its quadrupole moment
%of the effective jerk parameter
(see Appendix B of \citep{Heinesen:2020b}). 

Even though the maps in Figure~\ref{fig:200_params_skymap} are valid for a single observer, we see common signatures between observers\footnote{The particularly interested reader can find the equivalent of Figure~\ref{fig:200_params_skymap} for 100 different observers here:
\url{https://drive.google.com/drive/folders/1LIPmL05ENjLq4EnDBtLEyMAY2pRrcKm9?usp=sharing}
%\url{https://drive.google.com/drive/folders/1o-Q_rM0QE5LGkSioKFAPpRgmELUK_k9T?usp=sharing}
.}.
Specifically, the quadrupolar anisotropy tends to dominate $\mathfrak{H}$, the dipolar and octopolar signal typically dominates $\mathfrak{Q}$ and $\mathfrak{R}$, while the quadrupole and the 16-pole typically dominate the sky distribution of $\mathfrak{J}$. 
This is because, in general, spatial gradients of kinematic variables dominate the anisotropic effects over the observer's sky (with the exception of the spatial gradient of $a^\mu$ here, due to the small 4--acceleration amplitude of the observers). 
{While the amplitude of the anisotropies in $\mathfrak{H}$, $\mathfrak{Q}$, $\mathfrak{R}$, and $\mathfrak{J}$ are expected to vary with smoothing scale (and the associated density contrast), the  \emph{qualitative} anisotropic signatures of the effective cosmological parameters found in these simulations are expected to be robust to the choice of smoothing scale and/or the inclusion of a cosmological constant, as long as the observers are well described as comoving with the matter source in a general relativistic dust description. %This is because the terms in $\mathfrak{H}$, $\mathfrak{Q}$, $\mathfrak{R}$, and $\mathfrak{J}$ that involve the maximum number of spatial gradients of expansion rate and shear tend to dominate the anisotropic signal of the parameters in such space-times. 
%\hmr{we can maybe remove this last sentence - since you said this earlier in the paragraph!} \ahr{Ah yes, now removed.}
%We furthermore note, that even though the amplitude of the anisotropies as viewed over a typical observer's sky is dependent on the smoothing scale of the simulation, the anisotropic signature as viewed by an observer is largely uncorrelated with the density contrast at the observer's position within the given simulation.  
} 
%\ahr{This would be a good place to add the new figures!}\hmr{Thanks! I moved your paragraph on this down to after we have introduced $\Delta$, where I can introduce the figure too.}
\begin{figure*}%[h]
    %\centering
    \includegraphics[width=\textwidth]{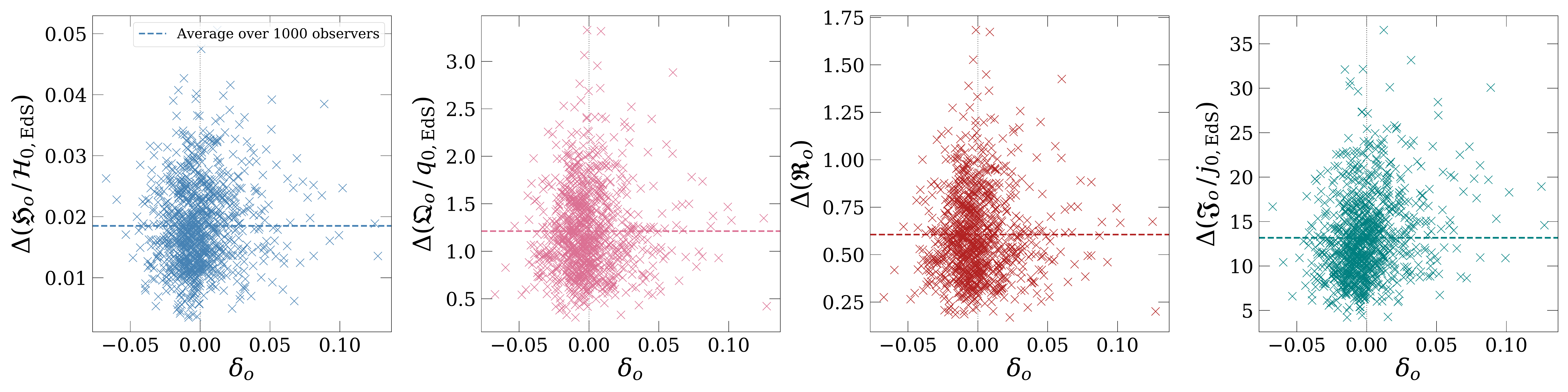}
    %{FLRW_pk_128_25p6Gpc_fullPk_maximalskyvar_effective_params_vs_EdS_eachobs_with_Obsavg_1x4_RandomSky_nobs1000_nLOS300_it648.pdf}
    \caption{Panels (left-to-right) show the maximal sky-variance \eqref{eq:deltaparams} for the effective Hubble, deceleration, curvature, and jerk parameters relative to their EdS counterparts. Points show $\Delta$ for 1000 observers in a simulation with a 200 $h^{-1}$ Mpc coarse-graining scale. Dashed horizontal lines show the average over all observers.
    }
    \label{fig:params_skyvariance}
\end{figure*}

% below is the OLD table -- MH21
% \begin{table*}
%     \centering
%     \begin{tabular}{|>{\centering}p{0.17\textwidth}| >{\centering}p{0.17\textwidth}| >{\centering}p{0.17\textwidth}| >{\centering}p{0.17\textwidth}| >{\centering\arraybackslash}p{0.17\textwidth}|}
%     \hline
%     \multirow{2}{4em}{ } & \multicolumn{2}{c|}{{\bf 100 $h^{-1}$ Mpc smoothing}} & \multicolumn{2}{c|}{{\bf 200 $h^{-1}$ Mpc smoothing}}\\\cline{2-3}\cline{4-5}
%     & Obs. mean & Obs. max & Obs. mean & Obs. max\\
%     \hline
%     ${\Delta}(\Eu_o/\mathcal{H}_{0,{\rm EdS}})$ & 0.091 & 0.19 & 0.067 & 0.15 \\
%     \hline
%     ${\Delta}(\mathfrak{Q}_o/q_{0,{\rm EdS}})$ & 18 & 53 & 5.5 & 14 \\\hline
%     ${\Delta}(\mathfrak{R}_o)$ & 9.01 & 26 & 2.8 & 7.2 \\\hline
%     ${\Delta}(\mathfrak{J}_o/j_{0,{\rm EdS}})$ & 790 & 4199 & 106 & 340 \\
%     \hline
%     \end{tabular}
%     \caption{Anisotropy of effective cosmological parameters across typical and extreme observers' skies. We show the mean and maximum sky-deviation $\Delta$ \eqref{eq:deltaparams} over 1000 observers in simulations with effective smoothing lengths of 100 and 200 $h^{-1}$ Mpc.}     \label{tab:params_skyvariances}
% \end{table*}
% below is the NEW table -- copied from erratum draft (sept 13th, 2021) (but with caption identical to MH21, sig figs. reduced)
\begin{table*}%[h!]
    \centering
    \begin{tabular}{|>{\centering}p{0.17\textwidth}| >{\centering}p{0.17\textwidth}| >{\centering}p{0.17\textwidth}| >{\centering}p{0.17\textwidth}| >{\centering\arraybackslash}p{0.17\textwidth}|}
    \hline
    \multirow{2}{4em}{ } & \multicolumn{2}{c|}{{\bf 100 $h^{-1}$ Mpc smoothing}} & \multicolumn{2}{c|}{{\bf 200 $h^{-1}$ Mpc smoothing}}\\\cline{2-3}\cline{4-5}
    & Obs. mean & Obs. max & Obs. mean & Obs. max\\
    \hline
    ${\Delta}(\Eu_o/\mathcal{H}_{0,{\rm EdS}})$ & 0.068 & 0.23 & 0.019 & 0.051 \\
    \hline
    ${\Delta}(\mathfrak{Q}_o/q_{0,{\rm EdS}})$ & 8.5 & 27 & 1.2 & 3.3 \\\hline
    ${\Delta}(\mathfrak{R}_o)$ & 4.3 & 13 & 0.61 & 1.7 \\\hline
    ${\Delta}(\mathfrak{J}_o/j_{0,{\rm EdS}})$ & 193 & 675 & 13 & 37 \\
    \hline
    \end{tabular}
    \caption{Anisotropy of effective cosmological parameters across typical and extreme observers' skies. We show the mean and maximum sky-deviation $\Delta$ \eqref{eq:deltaparams} over 1000 observers in simulations with effective smoothing lengths of 100 and 200 $h^{-1}$ Mpc.} \label{tab:params_skyvariances}
\end{table*}

We assess the level of anisotropy across an observer's sky by calculating the maximal sky-deviation $\Delta$ for each parameter
\citep[similar to][]{Andrade:2018}, e.g. for $\Eu$,
\begin{equation}\label{eq:deltaparams}
    \Delta(\Eu_o/\mathcal{H}_{0,{\rm EdS}}) \equiv \frac{\Eu_{o,{\rm max}} - \Eu_{o,{\rm min}}}{\mathcal{H}_{0,{\rm EdS}}},
\end{equation}
where $\Eu_{o,{\rm max}}$ is the maximum value of $\Eu$ across an observer's sky, and $\Eu_{o,{\rm min}}$ is the minimum.
Figure~\ref{fig:params_skyvariance} shows the maximal sky-variance $\Delta$ for the effective Hubble, deceleration, curvature, and jerk parameters, relative to their EdS counterparts (panels; left-to-right, respectively), as a function of the observers local density contrast, $\delta_o$. Points show $\Delta$ for 1000 observers, each with 300 \sayy{FullSky} lines of sight, placed in the simulation with a 200 $h^{-1}$ Mpc coarse-graining scale. Horizontal dashed lines show the average over all observers. Here we can see the anisotropic signature as viewed by an observer is largely uncorrelated with the density contrast at the observer's position.
%\hmt{Observers with small $\delta_o$}
%``average'' regions \aht{of vanishing density contrast} 
%measure large anisotropy. 
While the \emph{amplitude} of these anisotropic effects will depend on the smoothing scale (and therefore {the typical} density contrasts {in the simulation as a whole}){, observers with small $\delta_o$ can 
%{located in regions of small density contrast} 
measure the same level of anisotropy as observers
%located in regions 
with large $|\delta_o|$, \emph{within} a given model universe.} 
%\ahr{I reformulated the sentences above a bit, feel free to change back!}
%\hmr{I cut this a bit since we were being a little repetitive}
In Table~\ref{tab:params_skyvariances} we show the mean and maximum $\Delta$ across all observers shown in Figure~\ref{fig:params_skyvariance}, and for 1000 observers with the same lines of sight placed in the simulation with $100 h^{-1}$ Mpc smoothing length. 
%, each with 300 \sayy{FullSky} lines of sight, for both simulations with 100 and 200 $h^{-1}$ Mpc smoothing lengths.}
%the results shown in Figure~\ref{fig:params_skyvariance}.}
%show $\Delta$ for the observational effective parameters \eqref{eq:paramseff}. 
%Specifically, we show the mean and maximum $\Delta$ across all 1000 observers, each with 300 \sayy{FullSky} lines of sight.}
%, in simulations with 100 and 200 $h^{-1}$ Mpc smoothing lengths.
%\aht{We furthermore note, that even though the amplitude of the anisotropies as viewed over a typical observer's sky is dependent on the smoothing scale of the simulation, the anisotropic signature as viewed by an observer is largely uncorrelated with the density contrast at the observer's position within the given simulation.}

Analyses of the FLRW Hubble and/or deceleration parameters in SNe \citep{Kalus:2013,Bengaly:2015,Bengaly:2016,Andrade:2018,Colin:2018ghy} and galaxy data \citep{Bolejko:2015gmk,Migkas:2020fza}
come to diverse conclusions regarding the level of anisotropy in such data. 
These studies employ different phenomenological anisotropic modelling and %for 
consider different survey geometries. The truncation of the cosmographic representation and the relevant choice of smoothing scale will vary according to the redshift coverage of the survey in question (see Appendix~\ref{appx:series_convergence} for discussions on the level of approximation of the luminosity distance Taylor series expansion), 
and we thus expect different survey geometries to yield different empirical results on anisotropy in effective cosmological parameters. 
We should note that these works adopt FLRW cosmographic expansions -- as truncated according to FLRW model arguments -- in their search for anisotropy, which makes a direct comparison to our analysis difficult. 

Typical observers in our analysis have dipolar contribution to the deceleration parameter dominating over the monopolar contribution. 
%, as seen in Table~\ref{tab:params_skyvariances}
%\hmr{just to clarify: is this 'seen in Table 1' because the values of Delta for Q are large?} \ahr{Yes, but you are right that to fully see this we would need to show typical absolute values of the monopole as well, which will all be around $0.5$. Perhaps easiest solution is to erase 'as seen in Table~\ref{tab:params_skyvariances}'}.
This signature has interestingly been seen for us as observers in \citep{Colin:2018ghy} where the dipolar contribution to the deceleration parameter was found to dominate over the monopole out to scales of $z\sim 0.1$ in an empirical examination of SNe data.

%Further, truncating the series expansion at such low orders in redshift as in these works is problematic in the general anisotropic case, even while it may be valid in FLRW (see Appendix~\ref{appx:series_convergence}). 
%In addition, the physical interpretation of our chosen coarsegraining scales in the context of observational data is unclear. For these reasons, a meaningful comparison between these previous works adopting FLRW cosmography and our work in a general space-time framework is difficult.

\subsection{Sky averages of observational effective parameters}\label{sec:skyav}
%\subsubsection{Sky averages}
% =========================================================
% PLOTS of effective parameters rel to FLRW
% =========================================================
\begin{figure*}%[h]
    \centering
    \includegraphics[width=0.8\textwidth]{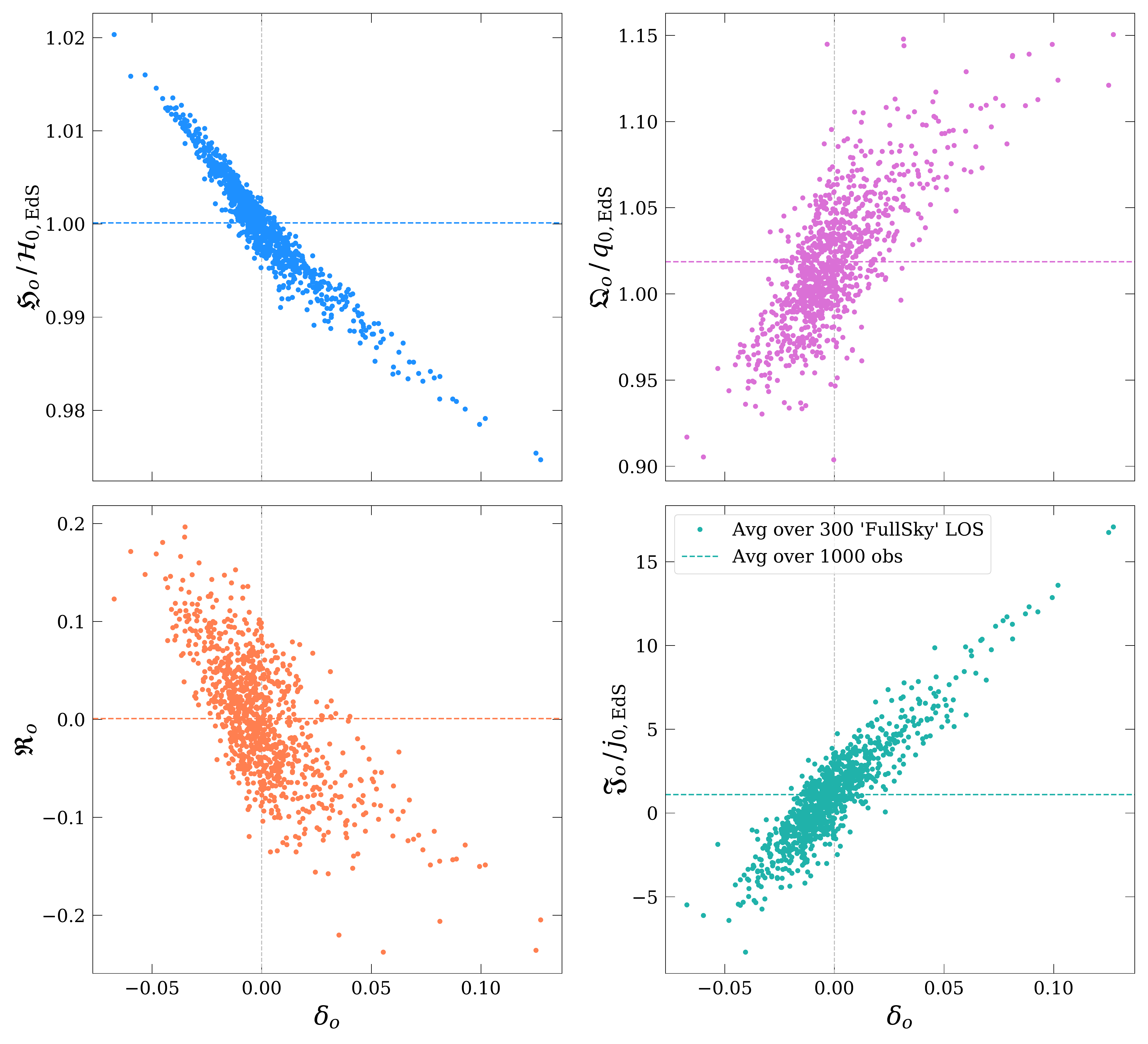}
    %{200Mpc_params_fullsky_128_difflims.pdf}
    \caption{Effective cosmological parameters \eqref{eq:paramseff} relative to their EdS counterparts (panels) calculated in an inhomogeneous numerical relativity simulation. Points show individual observers, with local density $\delta_o$, averaged over 300 randomly chosen lines of sight across their whole sky (see Figure~\ref{fig:fullsky}). Dashed lines of the same colour show averages over all points on each panel, and dot-dashed lines show \lcdm\, parameters relative to EdS.
    }
    \label{fig:200_params_full}
\end{figure*}

\begin{figure*}%[h]
    \centering
    \includegraphics[width=0.8\textwidth]{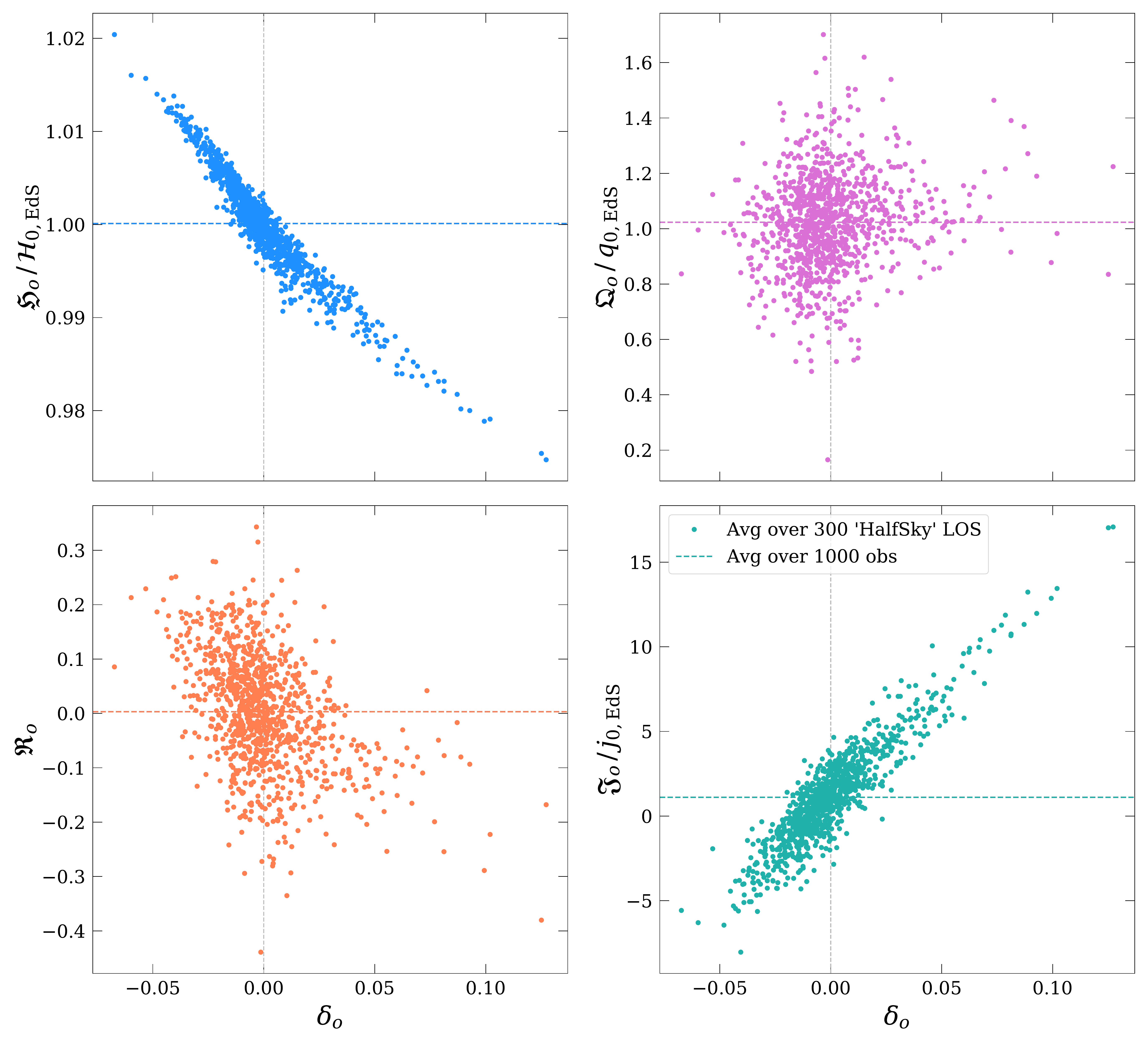}
    %{200Mpc_params_halfsky_128.pdf}
    \caption{Effective cosmological parameters \eqref{eq:paramseff} relative to their EdS counterparts (panels) calculated in an inhomogeneous numerical relativity simulation. Points show individual observers, with local density $\delta_o$, averaged over 300 randomly chosen lines of sight across \emph{half of} their whole sky (see Figure~\ref{fig:halfsky}). Dashed lines of the same colour show averages over all points on each panel, and dot-dashed lines show \lcdm\, parameters relative to EdS.
    }
    \label{fig:200_params_half}
\end{figure*}

Figures~\ref{fig:200_params_full} and \ref{fig:200_params_half} show our calculations of sky averages of the effective cosmological parameters in the simulation shown in Figure~\ref{fig:2Dslices} (200 $h^{-1}$ Mpc smoothing). Panels, top-left to bottom-right, show the effective observational Hubble, deceleration, curvature, and jerk parameters, respectively, each relative to their respective EdS parameter counterpart in \eqref{eq:dLFLRWcoef}. Points represent the effective parameters of individual observers as averaged over 300 lines of sight distributed according to a \sayy{FullSky} sample (Figure~\ref{fig:200_params_full}) and a \sayy{HalfSky} sample (Figure~\ref{fig:200_params_half}). See Figure~\ref{fig:skymaps} for examples of these sky samplings. Horizontal axes indicate the local density contrast $\delta_o\equiv \rho_o/\langle\rho\rangle_{\mathcal{D},{\rm all}}-1$ at each observer's position. % is \hmt{shown} on the horizontal axis of \hmt{each} panel. 
%In Figure~\ref{fig:200_params_full}, the observers have approximately isotropic lines of sight (\sayy{RandomSky} sampling), and in Figure~\ref{fig:200_params_half} observers have anisotropic lines of sight (\sayy{HalfSky} sampling). Figure~\ref{fig:skymaps} shows examples of each sky distribution. 
Dashed lines of the same colour as the points in each panel show the average over all 1000 observers (i.e., the average over all points), and dot-dashed black lines show the \lcdm\ value of the corresponding parameter (those panels without these lines are those where the \lcdm\ reference lies outside the limits of the plot). 

% below is the OLD table --- MH21
% \begin{table*}
%     \centering
%     \begin{tabular}{|>{\centering}p{0.17\textwidth}| >{\centering}p{0.17\textwidth}| >{\centering}p{0.17\textwidth}| >{\centering}p{0.17\textwidth}| >{\centering\arraybackslash}p{0.17\textwidth}|}
%     \hline
%     \multirow{2}{4em}{$\sigma({\rm sky\, avgs})$} & \multicolumn{2}{c|}{{\bf 100 $h^{-1}$ Mpc smoothing}} & \multicolumn{2}{c|}{{\bf 200 $h^{-1}$ Mpc smoothing}}\\\cline{2-3}\cline{4-5}
%     %\hline
%     & \sayy{FullSky} & \sayy{HalfSky} & \sayy{FullSky} & \sayy{HalfSky}\\
%     \hline
%     $\sigma(\Eu_o/\mathcal{H}_{0,{\rm EdS}})$ & 0.028 & 0.028 & 0.019 & 0.020 \\
%     \hline
%     $\sigma(\mathfrak{Q}_o/q_{0,{\rm EdS}})$ & 0.35 & 2.3 & 0.15 & 0.70
%     \\\hline
%     $\sigma(\mathfrak{R}_o)$ & 0.16 & 1.2 & 0.056 & 0.35\\\hline
%     $\sigma(\mathfrak{J}_o/j_{0,{\rm EdS}})$ & 170 & 164 & 23 & 23 \\
%     \hline
%     \end{tabular}
%     \caption{Cosmic variance of observational effective cosmological parameters \eqref{eq:paramseff}. We show standard deviations of the distribution of sky averages --- for the \sayy{FullSky} and \sayy{HalfSky} distributions (see Figure~\ref{fig:skymaps}) ---  for 1000 observers placed in simulations with effective smoothing lengths of 100 and 200 $h^{-1}$ Mpc.}
%     \label{tab:params_variances}
% \end{table*}
% below is the NEW table -- copied from the erratum overleaf sept 13th, 2021 (with caption matching MH21 exactly & sig figs reduced)
\begin{table*}%[h!]
    \centering
    \begin{tabular}{|>{\centering}p{0.17\textwidth}| >{\centering}p{0.17\textwidth}| >{\centering}p{0.17\textwidth}| >{\centering}p{0.17\textwidth}| >{\centering\arraybackslash}p{0.17\textwidth}|}
    \hline
    \multirow{2}{4em}{$\sigma({\rm sky\, avgs})$} & \multicolumn{2}{c|}{{\bf 100 $h^{-1}$ Mpc smoothing}} & \multicolumn{2}{c|}{{\bf 200 $h^{-1}$ Mpc smoothing}}\\\cline{2-3}\cline{4-5}
    %\hline
    & \sayy{FullSky} & \sayy{HalfSky} & \sayy{FullSky} & \sayy{HalfSky}\\
    \hline
    $\sigma(\Eu_o/\mathcal{H}_{0,{\rm EdS}})$ & 0.022 & 0.021 & 0.0057 & 0.0057 \\
    \hline
    $\sigma(\mathfrak{Q}_o/q_{0,{\rm EdS}})$ & 0.19 & 1.2 & 0.038 & 0.17
    \\\hline
    $\sigma(\mathfrak{R}_o)$ & 0.22 & 0.64 & 0.064 & 0.11 \\\hline
    $\sigma(\mathfrak{J}_o/j_{0,{\rm EdS}})$ & 43 & 43 & 2.9 & 2.9 \\
    \hline
    \end{tabular}
    \caption{Cosmic variance of observational effective cosmological parameters \eqref{eq:paramseff}. We show standard deviations of the distribution of sky averages --- for the \sayy{FullSky} and \sayy{HalfSky} distributions (see Figure~\ref{fig:skymaps}) ---  for 1000 observers placed in simulations with effective smoothing lengths of 100 and 200 $h^{-1}$ Mpc.}
    \label{tab:params_variances}
\end{table*}

In Table~\ref{tab:params_variances} we show the standard deviation for each distribution in Figures~\ref{fig:200_params_full} and \ref{fig:200_params_half}, as well as the same calculation in a simulation with a smaller coarse-graining scale of 100 $h^{-1}$ Mpc. The variances in the effective Hubble and jerk parameters show no appreciable change 
%increase by $\sim3-10\%$ 
moving from the \sayy{FullSky} to the \sayy{HalfSky} sampling, for both smoothing scales. However, we see drastic change in the effective deceleration and curvature parameters. For the deceleration, $\mathfrak{Q}_o$, we see an 
%$\sim 6.6\times$ ($\sim 4.7\times$)  <-- old sim
{$\sim 6.3\times$ ($\sim 4.5\times$)} % <-- new sim
increase in standard deviation when sampling only half of each observer's sky with a smoothing length of 100 (200) $h^{-1}$Mpc. The curvature parameter, $\mathfrak{R}_o$, shows a 
%$\sim7.5\times$ ($\sim6.2\times$) <-- old sim
{$\sim2.9\times$ ($\sim1.7\times$)} % <-- new sim
increase in standard deviation in the same case. 
%This similarity is to be expected, since the curvature parameter includes the deceleration parameter in its definition \aht{in (\ref{eq:paramseff})}.\hmr{I removed this sentence because we stated this in the previous paragraph}
These changes in variance can be visualised using the example observer sky map in Figure~\ref{fig:skymaps}. Consider averaging over our \sayy{HalfSky} distribution for the effective Hubble and jerk parameters for this observer. Since the quadrupolar mode is dominating the signal, cutting the sky in half (in the way we have done) should have minimal effect on the measured variance in those parameters. Considering either the effective deceleration or curvature parameter, cutting the sky in half \emph{should} affect the measured variance by this observer due to the dominance of the dipole.
%\hmr{clarify this last paragraph is specifically for our chosen sky sampling - unclear to PL}
%Since this change in variance is present considering the distribution of all observers supports our expectation that most observers share dominant anisotropic contributions.

Comparing Figures~\ref{fig:200_params_full} and \ref{fig:200_params_half} shows us that an observer can infer \emph{drastically} different effective cosmological parameters
when not fairly sampling their whole sky.
However, we emphasise that the variances presented here are valid only for our specific \sayy{HalfSky} distribution, and averages across different anisotropic sky-samplings will produce different variances. 
It is also important to note that the corrections to the EdS model expectation of the cosmological parameters decrease with an increase in smoothing scale. 
The choice of a physically relevant smoothing scale is therefore important and
%The relevant choice of smoothing scale in the physical modelling 
depends on the redshift span of the survey in question. 
%If we for
For instance, to consistently model a survey with minimum redshifts of $0.02$, corresponding to distances from the observer of $\sim 100$ $h^{-1}$ Mpc, we should not employ larger smoothing scales than $\sim 100$ $h^{-1}$ Mpc in our modelling. 
To ensure a consistent fit for the entire survey, the highest-redshift data
%On the other hand, the high redshift data points of the same survey 
should also be well approximated by the same cosmographic representation
%of the luminosity distance as applies to the data points at low redshifts
as the lowest-redshift data. %, for a consistent cosmological fit. 
In Appendix~\ref{appx:series_convergence}, we show that a consistent cosmographic representation of luminosity distance considering redshifts $\sim 0.02$ out to $z\sim1$ (and even $z\sim0.15$) is difficult in the presence of anisotropy. 
However, we stress that the interpretation of the observational effective cosmological parameters presented here remains valid as long as $z<1$ even if the truncated third order expansion \eqref{eq:series} breaks down as an accurate approximation of the exact function $z \mapsto d_L(z)$. 
%we discuss the level of approximation of the general series expansion of luminosity distance}
%We describe the expected convergence and level of approximation of the series expansion of luminosity distance \eqref{eq:series} in Appendix~\ref{appx:series_convergence} 
%in context of our simulations. 

We intend to investigate these issues relating to smoothing scale, and the impact of survey geometry on cosmological inference, in future work.

\subsection{Implications for a measurement on the local Hubble constant} \label{sec:H}

%\ahr{I think that we can here say that the isotropic component $\theta/3$ varies from observer to observer with deviance from the EdS background of typical values 2-3\% and up till till 5 percent. On top of this comes the anisotropic variation over the observers sky with the shear term which typically has amplitude of 5\% , but up till 8\% , cf. figure 1 (and figure 3 for example).  }\hmr{Isn't the \emph{total} isotropic + anisotropic variation in curlyh shown in the map in Fig3, for example? and fig 5. top left shows the total variation in curlyh, for an anisotropic sky... which is still only 2-3\% variance and up to 5?} 
%\ahr{Figure 3 only shows the anisotropic variation (The monopole theta/3 is constant over the sky for a single observer), but of course you plot it relative to $H_{EdS}$, so the mean tells you something about the monopole deviance from the EdS value of this particular observer. The current figure 3 has less variance over the sky than the previous, so only 2 percent shear effect in this case. Figure 5 shows the variation between observers for the half sky average, but half sky averages tend to exclude the quadrupolar signal, and so for other survey geometries we would expect more variation. Potentially go back to your old plots of individual rays (no sky averaging) to see the entire magnitudes of the effects? We should get around 10\% in combined effects for the most extreme lines, even though we would have to be special observers with special survey geometry to see this.}

The sky averages shown in Figure~\ref{fig:200_params_full} are representative of the monopole (isotropic) contribution to the effective parameters. Focusing on the effective Hubble parameter (top left panel, with variances in the top row of Table~\ref{tab:params_variances}), we find isotropic variances of 
%2-3\%  <-- old sim
{0.5\% (2\%)} %<-- new sim
with respect to the EdS value, with maximum variances of up to 
%5\% (7.5\%) <-- old sim
{2\% (6\%)} %<-- new sim
on scales 200 (100) $h^{-1}$ Mpc. This represents the variance in the $\theta/3$ term in \eqref{def:Eevolution} when moving between different positions in the simulation. 
Figure~\ref{fig:200_params_full} shows a clear and physically expected correlation between values of $\theta/3$ and the local density. 
%We expect local fluctuations in the density field to be the main cause of fluctuations in $\theta$. 
For similar local density contrasts $\delta$ (and/or similar smoothing scales), we find the above variance is in broad agreement with cosmic variance in the FLRW local Hubble parameter studied analytically \citep{Marra:2013,Camarena:2018} and in the context of Newtonian N-body \citep{Wotjak:2014,Odderskov:2014,WuHuterer:2017} and NR simulations \citep{Macpherson:2018akp}. 

%This variance is in broad agreement with the $\sim1-3\%$ cosmic variance in the Hubble parameter found in the context of analytic studies \citep[e.g.][]{Marra:2013}, and the $\sim 1\%$ found in Newtonian N-body {WuHuterer:2017} and NR simulations \citep{Macpherson:2018akp} (we note that the latter contains larger variations up to 5\% when considering the same scales studied here) \ahr{Yes, comparing results for the same scales is important here. State scale and corresponding density contrasts of these studies? The density contrast is actually probably the most important for the variance in theta.}. %We note a slight increase in variance with respect to previous simulation results, most likely due to differences in defining the \sayy{smoothing scale} of such analysis. For example, in \citet{Macpherson:2018akp}, variance in an effective Hubble parameter was smoothed over the \emph{entire} survey volume to give a final estimated $\sim 1\%$ variance on a local measurement of $H_0$.

In addition to this variance based on inhomogeneity (observer position), we have an  \emph{anisotropic} contribution to the effective Hubble parameter $\Eu$, which will vary across each observer's sky. In this work, the anisotropy in $\Eu$ comes primarily from the shear tensor, i.e. the third term in \eqref{def:Eevolution}. 
The top row of Table~\ref{tab:params_skyvariances} 
%quantifies the mean and maximum sky-variance in $\Eu$ for all observers studied here. 
shows that typical observers measure 
%$9.1\%$ ($6.7\%$)  <-- old sim
$6.8\%$ ($1.9\%$) % <-- new sim
maximum deviation in $\Eu$ across their sky for 100 (200) $h^{-1}$ Mpc coarse-graining scales. %\hmr{None of our observers in the 200 smoothing scale measure $>10\%$ variance in $\Eu$, so I changed the following sentence slightly:} 
{We also find that 11\% of the observers measure >10\% maximal deviation $\Delta$ in $\Eu$ across their sky in the simulation with 100 $h^{-1}$ Mpc coarse-graining scale.}
%We also find that 37\% (7.6\%) of the observers measure >10\% maximal deviation $\Delta$ in $\Eu$ across their sky.

However, this is not necessarily indicative of how many observers will measure a \emph{sky-average} of $\Eu$ to be $\sim 10\%$ larger than the global mean \citep[as found in, e.g.,][for the local FLRW Hubble parameter]{Riess:2019cxk}. The specific survey geometry will affect the number of observers we find to measure a higher local $\Eu$.
We intend to investigate the role of anisotropy in context of the Hubble tension in detail in future work.

\section{Conclusions}\label{sec:conclude}

The general luminosity-distance redshift relation presented by \citet{Heinesen:2020b} offers the potential to completely relax the assumptions of exact homogeneity and isotropy at the base of most cosmological data analysis. 
%for a completely model-independent analysis of cosmological data. 
Additional degrees of freedom introduced --- because of the lack of assumptions made --- means that more cosmological data is required in order to use this framework to convey \emph{fully} model-independent data analysis.

We have calculated the observational effective cosmological parameters \eqref{eq:paramseff} in simulations with realistic initial conditions evolved with numerical relativity. Our simulations therefore share the qualities of the general formalism in that they contain no assumptions of a global background metric. 
%whatsoever
We have used conservative coarse-graining scales to study the variance of these parameters on scales where the simulated model universe is well within the linear regime of density contrasts. 
%relative to an EdS background model}. 
%We find significant anisotropy in the 
We find that effective cosmological parameters can be significantly anisotropic across the observers' skies (see Figure~\ref{fig:skymaps} for an example), with corrections to the relevant FLRW parameters even in the monopole limit of a fairly sampled sky.

Considering a cosmographic representation of luminosity distance with a 200 $h^{-1}$ Mpc smoothing scale,
%\hmr{Your sentence here is new - we haven't referred to the smoothing scale being interpreted in this way (and so probably shouldn't state it in the conclusions), so we can either add it somewhere in an earlier section or leave it out?} \ahr{I decided to just leave it out for simplicity!}
our main conclusions are: % \hmr{suggestion: also add, or replace these, with the conclusions from the 100 Mpc/h smoothing? they are a bit more interesting now...}
\begin{itemize}
    \item The effective Hubble parameter has 
    {$0.5\%$} cosmic variance between observers relative to the EdS value in the monopole limit, i.e. a fairly-sampled sky. We find that {$\sim$7\%} of observers measure a Hubble parameter {$>$3\%} different from the EdS value in this limit (see Table~\ref{tab:params_variances} and Figure~\ref{fig:200_params_full}).
    
    \item Maximal quadrupolar anisotropies in the effective Hubble parameter across an  observers sky are typically {2\%}, and can be as large as {5\%} (see Table~\ref{tab:params_skyvariances}).
    
    \item A uniform sky-average of the effective deceleration parameter has 
    standard deviation of {4\%} between observers relative to the EdS value (see Table~\ref{tab:params_variances}).
    
    \item The dipolar signal of the effective deceleration parameter dominates the monopolar contribution for typical observers, with the mean observer seeing a {120\%} deviation between highest and lowest value
    on their sky (see Table~\ref{tab:params_skyvariances}).
    
    \item A half-sampled sky can bias some observers measurements such that they measure acceleration \emph{without} any actual accelerated expansion of space.
\end{itemize}

As a final note, since our study is concerned only with large scales --- and therefore small density contrasts --- we expect our results to be well-approximated in the weak-field limit of general relativity.
%\hmt{As a final note, even though our study is concerned only with large scales --- and therefore small density contrasts --- the main contribution to the anisotropic effects we find are caused by \emph{derivatives} of local kinematic variables, which may be large even if density contrasts are small. We therefore expect linear perturbation theory to be insufficient to capture all relevant effects.However, we do expect our results to be well-approximated in the weak-field limit of general relativity.} 
%\ahr{I am not sure about the above statement, as for the purpose of solving Einsteins equations (containing up to second derivatives of the metric) in a linearised way, we only need to assume that fluctuations in the metric and its gradients up till second order are small (so that density contrasts but not neccesarily their derivatives are small)?. I would therefore formulate a bit more vaguely such statements.}
Therefore, a re-analysis of the general observational effective parameters in the context of 
%Newtonian N-body, 
weak-field N-body cosmological simulations \citep[i.e. \texttt{gevolution}, see][]{Adamek:2015eda,Adamek:2016,Adamek:2018rru}, or simulations using the fully-constrained formulation of GR \citep[i.e. \texttt{GRAMSES}, see][]{Barrera-Hinojosa:2019mzo,Barrera-Hinojosa:2020arz} %cosmological simulations 
may produce similar results.
This could also
%for instance 
be investigated in the context of constrained cosmological simulations reconstructing the environment surrounding the local group, as done in, e.g., \citep{Carlesi:2016qqp}.
%\ahr{I am not sure about the following sentence, since it seems to suggest that predictions can be improved by using Newtonian codes relative to your simulations? \hmr{yes, I guess that's what I meant - if we do see similar results in the Newtonian case, then the much higher-resolution Newtonian codes available could produce more precise results, but you're right that it is ambiguous so I'll remove it}
%One thing that could be mentioned is that cosmological simulations which are designed for reproducing the local cosmic environment of us as observers }
%Such an analysis in higher-resolution simulations is promising for making precise predictions for upcoming surveys \aht{within the framework of quasi-Newtonian cosmology}. 

%We have restricted our study to large scales and therefore small density contrasts with maximum $|\delta|\sim0.1-0.2$. Our simulations should therefore be well-approximated by the weak-field limit of general relativity. 
% %Nonetheless, even in this regime we find appreciable deviation from EdS parameters. 
% It would be interesting to re-analyse the observational effective parameters \eqref{eq:paramseff} in the context of Newtonian N-body simulations.

%\hmr{The following sentence is too long, break up} 
Our main conclusions %of our analysis
suggest that the consideration of local anisotropies could be important for cosmological analysis. 
%\hmt{The consistent, anisotropic} 
The anisotropic cosmographic representation of luminosity distance
%which allows for anisotropies 
\citep{Heinesen:2020b} used in this work gives us a framework to interpret near-future large cosmological surveys in a completely model-independent way. 
%to interpret low-redshift data is necessary to 
This may be necessary to ensure we draw correct conclusions about the cosmic expansion and acceleration 
%expansion degrees of freedom, cosmic acceleration, etc., 
of the Universe. % from \hmt{precision} data.

%\subsection{Caveats}

%\hmr{Outline the main caveats of the simulations / calculations, and room for improvement}

%\hmr{Main caveats of my simulations are: 1) fluid approx, i.e. no virialisation, 2) no lambda, matter dominated and therefore structure looks different, 3) periodic BCs and potential for global R-->0, 4) modes sampled to ~ 2 grid cells, and so may be under-resolved. But I don't think any of those are an issue here... except maybe the last one, but we found similar results in the case with larger, smoothed structures.}

% Figure template
% \begin{figure}[h]
%     \centering
%     \includegraphics[width=0.7\columnwidth]{filename}
%     \caption{ 
%     }
%     \label{fig:tmpfig}
% \end{figure}

\acknowledgments

The authors would like to thank {Julian Adamek,} Thomas Buchert, Ruth Durrer, Martin France, and Paul Lasky for helpful comments on the manuscript. The authors would also like to thank the anonymous referees whose comments improved the quality and clarity of the manuscript. HJM appreciates support received from the Herchel Smith Postdoctoral Fellowship Fund. This work is part of a project that has received funding from the European Research Council (ERC) under the European Union's Horizon 2020 research and innovation programme (grant agreement ERC advanced grant 740021--ARTHUS, PI: Thomas Buchert).
%\hmr{and others? e.g. other useful emails we've gotten?}} \ahr{I added Thomas since he did a few comments as well. We can move this to the top, such that we thank named people before anonymous referees?}
This work used the DiRAC@Durham facility managed by the Institute for Computational Cosmology on behalf of the STFC DiRAC HPC Facility (www.dirac.ac.uk). The equipment was funded by BEIS capital funding via STFC capital grants ST/P002293/1, ST/R002371/1 and ST/S002502/1, Durham University and STFC operations grant ST/R000832/1. DiRAC is part of the National e-Infrastructure.

\appendix

%%%%%%%%%%%%%%%%%%%%%%%%%%%%%%%%%%%%%%%%%%%%%%%%%%%%%%%%%%%%%%%%%%%%%%%%%%%%%%%%%%%
\section{Convergence of the series expansion}\label{appx:series_convergence}
%%%%%%%%%%%%%%%%%%%%%%%%%%%%%%%%%%%%%%%%%%%%%%%%%%%%%%%%%%%%%%%%%%%%%%%%%%%%%%%%%%%

We now examine the convergence and quality of the approximation of the Taylor series (\ref{eq:dLexpand2}). 
For investigating these properties of the series expansion in detail, we must ideally employ ray tracing algorithms for comparison to the exact expression. 
Employing ray tracing is indeed our 
%long term 
goal for future work, but for the purpose of analysing the results of this paper, we shall rely on crude order of magnitude arguments. 
We can note that the FLRW series expansion of luminosity distance in redshift is convergent for redshifts of $z < 1$, after which the series in general breaks down \citep{Cattoen:2007sk}. Following the arguments in Section 4.1 of \citep{Cattoen:2007sk}, we might define $\mathfrak{H}_o$, $\mathfrak{Q}_o$ and $\mathfrak{J}_o$ through the series expansion of $1/(1+z) \equiv \frac{E_o}{E}$ around the point of observation in the following way:  
\begin{equation}
\begin{aligned}
\label{eq:zexpand}
     \hspace*{-0.1cm} \frac{1}{1+z}  & =  1 +   \mathfrak{H}_o  E_o \Delta \lambda - \frac{1}{2} ( \mathfrak{Q}_o + 1) \mathfrak{H}^2_o E^2_o \Delta \lambda^2  \\ 
    & + \frac{1}{6}(\mathfrak{J}_o + 4 \mathfrak{Q}_o + 3) \mathfrak{H}^3_o E^3_o \Delta \lambda^3 + \mathcal{O}(\Delta \lambda^4)  \, ,  
\end{aligned}
\end{equation}
where $\Delta \lambda \equiv \lambda - \lambda_o$, and where $\lambda$ is an affine parameter of the null ray. 
We see that the series expansion (\ref{eq:zexpand}) has a pole at $z=-1$ for each null ray, and the radius of convergence must thus be at most $|z|=1$, such that the series also fails to converge at $z > 1$. 
Consequently, when inverting this series, to obtain $\Delta \lambda$ as a function of $z$, we should not expect this series to be convergent for $z>1$ either. 
Thus, we do not expect (\ref{eq:dLexpand2}) to converge beyond $z>1$, but have no reason to believe that it will diverge at smaller scales either, if the regularity requirements discussed in \citep{Heinesen:2020b} are satisfied in addition (which is the case in our analysis). 

The quality of the approximation of the Taylor series (\ref{eq:dLexpand2}) truncated at third order can however be very poor at redshifts approaching 1, and we expect this to be the case in the present simulation setup for most observers for the following reason. 
%The reason is that 
Spatial gradients of order ($n-1$) of kinematic fluid variables enter in the $n$'th coefficient $d_L^{(n)}$ in (\ref{eq:dLexpand2}) -- see for instance $\overset{1}{\mathfrak{q}}_\mu$ and $\overset{3}{\mathfrak{q}}_{\mu \nu \rho }$ in \eqref{qpoles} which contain the first order spatial gradient of the expansion rate and of the shear tensor, respectively. 
We can make the following order of magnitude estimate of the $n$'th term $d_L^{(n)}z^n$ in the Taylor series (\ref{eq:dLexpand2}) in terms of the smoothing scale $\Delta X$ of the simulation and $\mathfrak{H}_o$ 
\begin{equation}\label{eq:order}
\hspace*{-0.2cm}	\left| \frac{d_L^{(n)} z^n }{d_L^{(1)} z} \right| \lesssim \frac{1}{n!} \frac{z^{n-1}}{(\Delta X \mathfrak{H}_o )^{n-1}} {\rm sup}\left( \left|  \frac{\Delta \mathfrak{H}_o }{\mathfrak{H}_o} \right| \right) \, , \quad n\geq 2
\end{equation}
where $\Delta \mathfrak{H}_o$ is the increase of $\mathfrak{H}_o$ to a neighbouring grid cell of the observer, and ${\rm sup}$ denotes the supremum (here corresponding to the maximum value) over all grid cells. 
%Plugging in
Substituting the EdS value of the Hubble constant $\mathcal{H}_{\rm 0, EdS}\approx 45$ km/s/Mpc for $\mathfrak{H}_o$, gives $1/(\Delta X \mathfrak{H}_o) \sim 50 \times 100 \, h^{-1} {\rm  Mpc} / \Delta X$. 

Let us first consider $\Delta X =100 \, h^{-1}$Mpc for which we use the estimate ${\rm sup}\left( \left|  \Delta \mathfrak{H}_o / \mathfrak{H}_o \right|  \right) \sim 0.1$ which corresponds to $\sim 3\times \sigma(\Eu/\mathcal{H}_{0,{\rm EdS}})$ given in Table~\ref{tab:params_variances}. With these values we have $\left| d_L^{(m)} / d_L^{(1)} \times z^{m-1} \right|  \lesssim 0.01$ for all $m \geq 4$ when $z\lesssim 0.03$. The coarse-graining scale of $\Delta X =100 \, h^{-1}$Mpc itself corresponds to $z \sim 0.02$. {We therefore expect the Taylor series expansion as truncated at third order to approximate the exact luminosity distance to within one percent for $0.02\lesssim z \lesssim 0.03$. }   
%of the observers
%to within one percent for $0.02\lesssim z \lesssim 0.03$ %, and we thus do not expect scales below this in redshift to be accurately modelled in this description. %We therefore expect the Taylor series expansion truncated at third order to approximate the exact luminosity distance
%of the observers
%to within one percent for $0.02\lesssim z \lesssim 0.03$. 

Let us next consider $\Delta X =200 \, h^{-1}$Mpc. We have a similar order of magnitude estimate of ${\rm sup}\left( \left|  \Delta \mathfrak{H}_o / \mathfrak{H}_o \right|  \right) \sim 0.1$ for these scales. 
%, corresponding to $\sim 3\times \sigma(\Eu/\mathcal{H}_{0,{\rm EdS}})$ given in Table~\ref{tab:params_variances}. 
In this case we satisfy $\left| d_L^{(m)} / d_L^{(1)} \times z^{m-1} \right|  \lesssim 0.01$ for all $m \geq 4$ when $z\lesssim 0.06$. 
The coarse-graining scale of $\Delta X =200 \, h^{-1}$Mpc corresponds to $z \sim 0.04$, and we expect the truncated Taylor series \eqref{eq:dLexpand2} to provide an approximate luminosity distance to within one percent for $0.04\lesssim z \lesssim 0.06$. 

If we for instance wanted to approximate scales out to $z = 0.15$, the minimum integer, $m_{\rm max}$, satisfying $\left| d_L^{(m)} / d_L^{(1)} \times z^{m-1} \right|  \lesssim 0.01$ for all $m \geq m_{\rm max}$ is $m_{\rm max} = 9$ (18) 
%($m_{\rm max} = 18$) 
for a smoothing scale 
%$\Delta X =200 \, h^{-1}$Mpc ($\Delta X =100 \, h^{-1}$Mpc).
$\Delta X = 200$ (100) $h^{-1}$Mpc 
Thus, in order to obtain a consistent cosmography valid from scales of $200 \, h^{-1}$Mpc ($z\approx0.04$) and out to $z=0.15$, with error terms of $\lesssim 1\%$, we expect to have to include terms up to $8^{\rm th}$ order in the series expansion of $d_L$. 

We note that as long as we remain within the radius of convergence of the Taylor series expansion of $d_L$ in $z$, the interpretation of the parameters $\{\Eu,\mathfrak{Q},\mathfrak{J},\mathfrak{R}\}$ as effective observational Hubble, deceleration, curvature and jerk parameters is preserved. 
While the radius of convergence must be examined in detail with ray tracing codes, we \textit{a priori} have no reason to expect failure of convergence before $|z|=1$ is reached, as per the discussion above.

It might seem surprising that the cosmographic representation of luminosity distance \eqref{eq:dLexpand2} is not perturbatively close the analogous expression for the background EdS model, since the smoothing scales employed in the present analysis are such that density contrasts are within the linear regime. However, the higher order effective cosmological parameters $\{\mathfrak{Q},\mathfrak{J},\mathfrak{R}\}$ can be non-linear even if the density field is in the linear regime on account of the spatial gradients entering the various multipole coefficients (see for instance the multipole expansion of $\mathfrak{Q}$ in \eqref{qpoles}). 
Since our analysis is in the linear density contrast regime, we expect our results to be repeatable in Newtonian simulations. 

We could of course have employed even larger coarse-graining scales, for a well behaved Taylor series out to larger $z$. However, this would come at the price of a poorly approximated distance-redshift relation at small scales. %\hmr{perhaps move our motivation here for choosing 100 Mpc as this corresponds to the minimum scale sampled in local H0 analysis?}
Our analysis suggests that a well-defined, collective cosmographic representation of distance-redshift data from scales $\sim 100 h^{-1}$ Mpc and up to redshifts of $z\sim 1$ is challenging.

\section{\texttt{Mescaline} calculations}\label{appx:mesc}
%%%%%%%%%%%%%%%%%%%%%%%%%%%%%%%%%%%%%%%%%%%%%%%%%%%%%%%%%%%%%%%%%%%%%%%%%%%%%%%%%%%

\texttt{Mescaline} is a post-processing analysis code for Einstein Toolkit 
%HDF5 
data \citep{Macpherson:2019a}. Here we use an extended version of \mesc\, to calculate each of the terms in the series expansion \eqref{eq:series} for a chosen number of observers (randomly placed in the simulation domain) each with a chosen number of lines of sight $e^\mu$ (either randomly chosen across the whole sky or within a restricted region). 

\texttt{Mescaline} contains no physical assumptions on the form of the metric, fluid, or extrinsic curvature. 
%\ahr{Moved this paragraph up from the bottom.}
\texttt{Mescaline} reads data output from the Einstein Toolkit, in Hierarchical Data Format 5 (HDF5), and is therefore written in terms of $3+1$ variables associated with the coordinate representation (\ref{eq:ds}) of the general metric. Specifically, it reads the spatial metric $\gam_{ij}$, the extrinsic curvature $K_{ij} \propto \pd_t \gam_{ij}$, the lapse function $\alpha$, the rest-mass density $\rho$, and the 3-velocity $v^i$ of the fluid (with respect to the Eulerian observer comoving with the foliation defined by $t$) for the whole spatial grid for all relevant time steps.  
The code enforces a gauge condition with shift vector $\beta^i=0$ (but for any $\alp$ and $\pd_t \alp$), and a uniform Cartesian grid with periodic boundary conditions (equivalent to applying a torus condition on the topology of the spatial sections)%, and geometric units with $G=c=1$. \ahr{Erase last part with the geometric units. The reader knows from the introduction section that we are providing quantities in these units. } 
All derivatives (in time and space) are taken using fourth-order finite differences. 

In this appendix we outline the key calculations required for each of the terms \eqref{eq:dLexpand2}. To calculate the effective Hubble parameter, $\mathfrak{H}$, we need the volume expansion rate, shear tensor and 4-acceleration, respectively defined as
\begin{equation}\label{eq:fluid_vars}
\begin{aligned}
    \theta &\equiv \nabla_\mu u^\mu, \\
    \sigma_{\mu\nu} &\equiv b^{\alpha}_{\mu}\, b^{\beta}_{\nu}\, \nabla_{(\alpha} u_{\beta)} - \frac{1}{3}\theta \, b_{\mu\nu}, \\
    a^\mu &\equiv u^\nu \nabla_\nu u^\mu.
\end{aligned}
\end{equation}
Here, $u^\mu$ is the 4-velocity of the fluid (which we choose to coincide with the 4-velocity of the observers), round brackets imply symmetrisation over indices, $\nabla_\mu$ is the Levi-Civita connection (covariant derivative) associated with $g_{\mu\nu}$, and $b_{\mu\nu} \equiv g_{\mu\nu} + u_\mu u_\nu$ is the spatial projection tensor in the frame of the fluid flow.

\subsection{Derivatives along the null ray}

To calculate the effective cosmological parameters \eqref{eq:paramseff}, we need the first and second derivatives of $\mathfrak{H}$ along the null ray, i.e., the derivatives with respect to affine parameter $\lambda$. In terms of $3+1$ variables, these are
\begin{align}\label{eq:dhds}
	\frac{{\rm d}\mathfrak{H}}{{\rm d}\lambda} &= k^\mu \nabla_\mu \mathfrak{H}, \\
	&= k^0 \pd_t \mathfrak{H} + k^i \pd_i \mathfrak{H},
\end{align}
and therefore
\begin{equation}\label{eq:d2hds2}
	\begin{aligned}
		\dtwoh =& (k^0)^2 \pd^2_t \mathfrak{H} + 2 k^0 k^i \pd_t \pd_i \mathfrak{H} + k^i k^j \pd_i \pd_j \mathfrak{H} \\
		& - \frac{\pd_t \mathfrak{H}}{\alp} \left[ (k^0)^2\pd_t\alp + 2k^0k^i\pd_i \alp - k^i k^j K_{ij} \right] \\
		& - \pd_i \mathfrak{H} \left[ (k^0)^2 \alp \gam^{ij} \pd_j \alp - 2\alp k^0 k^j K^i_{j} + k^j k^k \Gam^i_{jk}\right],
	\end{aligned}
\end{equation}
where $t$ is the coordinate time of the simulation, $\pd_\mu \equiv \pd/\pd x^\mu$, and $\Gam^i_{jk}$ are the Christoffel symbols associated with the spatial metric $\gam_{ij}$. In deriving \eqref{eq:d2hds2}, we have used the time components of the 4-Christoffel symbols associated with the metric $g_{\mu\nu}$, namely,
\begin{subequations}
\begin{align}
	{}^{(4)}\Gamma^0_{00} &= \frac{1}{\alp}\pd_t\alp, \quad \quad {}^{(4)}\Gamma^i_{j0} = - \alp K^i_{j}, \\
	{}^{(4)}\Gamma^i_{00} &= \alp \gam^{ij}\pd_j \alp, \quad {}^{(4)}\Gamma^0_{0i} = \frac{1}{\alp}\pd_i \alp, \\
	{}^{(4)}\Gamma^0_{ij} &= -\frac{1}{\alp}K_{ij},
\end{align} 
\end{subequations}
and we note that for $\beta^i=0$ we have ${}^{(4)}\Gamma^i_{jk} = \Gamma^i_{jk}$. 

\subsection{4-Ricci tensor}

For the effective curvature parameter, $\mathfrak{R}$, we need the components of the (symmetric) 4-Ricci tensor, $R_{\mu\nu}$. In terms of $3+1$ variables, the time-time component is
\begin{align}
	R_{00} = \alp \pd_j (\alp) &\pd_i \gam^{ij} + \alp \gam^{ij} \pd_i \pd_j \alp + \alp \pd_t K \\ 
	&+ \alp \Gam^i_{ji} \gam^{jk}\pd_k \alp - \alp^2 K_{ij}K^{ij},\nonumber
\end{align}
the time-space components are
\begin{align}
	R_{0i} & = \alp \pd_i K - \alp \pd_j K^j_{i} - \alp \Gam^j_{kj} K^k_{i} + \alp \Gam^j_{ki} K^k_{j},
\end{align}
and the spatial components are
\begin{align}
	R_{ij} = {}^{(3)}\mathcal{R}_{ij} &+ K K_{ij} - 2 K^l_{j} K_{il} \\
	&- \frac{1}{\alp} \left( \pd_t K_{ij} + \pd_i \pd_j \alp - \pd_k (\alp) \Gam^k_{ij} \right), \nonumber
\end{align}
where ${}^{(3)}\mathcal{R}_{ij}$ is the Ricci tensor of the spatial surfaces, and $K=K^i_{\phantom{i}i}$.

\subsection{Direction vector}\label{sec:emu}

%\ahr{I rewrote this section slightly, changing the order of how things are mentioned. See if you agree with the logic!} \hmr{looks good!}

The photon 4--momentum, $k^\mu$, of an incoming null ray can always be decomposed based on the observer 4--velocity, $u^\mu$, in the following way:  
\begin{equation}
    k^\mu = E(u^\mu - e^\mu),
\end{equation}
where $E\equiv -k^\mu u_\mu$ is the observed energy of the photon, and $e^\mu$ denotes the spatial direction of observation satisfying the orthonormal requirement 
\begin{equation}\label{eq:emuconstraints}
	g_{\mu\nu} e^\mu e^\nu = 1, \quad e^\mu u_\mu = 0.
\end{equation} 
The direction vector $e^\mu$ specifies the photon 4--momentum uniquely up to the normalisation $E$. 
We generate sky-maps of lines of sight of the observers in our analysis (see Figure~\ref{fig:skymaps} for example sky-maps) by setting 
\begin{equation}
	e^i = D \, m^i , 
\end{equation}
where $m^i$ are pseudo-random numbers. To generate the so-called \sayy{FullSky} distribution (see Figure~\ref{fig:fullsky}) for each observer, each component of $m^i$ is drawn from a uniform distribution over the range $[-1,1)$. To generate the \sayy{HalfSky} distribution (see Figure~\ref{fig:halfsky}), two of the components of $m^i$ are drawn from a uniform distribution over the interval $[-1,1)$, while the remaining component is constrained to the range $[0,1)$. % \ahr{correct?}. \hmr{yep!}
The normalisation factor $D$ and the time-component\footnote{The coordinate system $\{t,x^i\}$ used in the simulations is in general not adapted to the fluid flow, and we will thus in general have $e^0 \neq 0$.} $e^0$ are determined from the constraints \eqref{eq:emuconstraints}, which give  
\begin{align}
	e^0 &= \frac{\pm m^i u_i}{\sqrt{u_0^2 \gam_{ij} m^i m^j - \alp^2 m^i m^j u_i u_j}}, \\
	D &= \frac{\mp u_0}{\sqrt{u_0^2 \gam_{ij} m^i m^j - \alp^2 m^i m^j u_i u_j}}.
\end{align}

%\hmt{The effective cosmological parameters are anisotropic, and therefore require %specification of a direction of observation, $e^\mu$. In addition, to take derivatives %along the null ray we require the photon 4--momentum $k^\mu$.}
%The direction of observation is a unit-vector field orthogonal to $u^\mu$, therefore %satisfying
%\begin{equation}\label{eq:emuconstraints}
%	g_{\mu\nu} e^\mu e^\nu = 1, \quad e^\mu u_\mu = 0.
%\end{equation}
%While $e^\mu$ is a \emph{spatial} vector (i.e. $e^0=0$) in the observers frame, the simulations we use here are performed in a general frame (\emph{not} comoving with the fluid flow), and so in this general coordinate system we will have $e^0 \neq 0$. %Therefore, along with the constraints \eqref{eq:emuconstraints}, we have two degrees of freedom to specify $e^\mu$. 
%For each individual line of sight (defined by a unique $e^\mu$) we choose
%\begin{equation}
%	e^i = D \, m^i
%\end{equation}
%where $m^i$ are pseudo-randomly drawn numbers in the range $[-1,1)$, and $D$ is a factor to be determined. Solving the constraints \eqref{eq:emuconstraints} gives
%\begin{align}
%	e^0 &= \frac{\pm m^i u_i}{\sqrt{u_0^2 \gam_{ij} m^i m^j - \alp^2 m^i m^j u_i u_j}}, \\
%	D &= \frac{\mp u_0}{\sqrt{u_0^2 \gam_{ij} m^i m^j - \alp^2 m^i m^j u_i u_j}}.
%\end{align}
%The photon 4--momentum is then calculated using the decomposition
%\begin{equation}
%    k^\mu = E(u^\mu - e^\mu),
%\end{equation}
%where $E\equiv -k^\mu u_\mu$ is the observed energy of the photon. 

\subsection{Analytic test}\label{sec:analytic_test} 
The calculations of $\mathfrak{H}$ and its derivatives \eqref{eq:dhds} and \eqref{eq:d2hds2}, as presented above are new in \mesc.
%Some of the calculations presented in this Appendix are new to \mesc, specifically $\mathfrak{H}$ and its derivatives \eqref{eq:dhds} and \eqref{eq:d2hds2}. 
We must therefore test these calculations against a suitable analytic solution to ensure they are accurate, with errors converging at the expected rate.

Rather than passing in simulation output HDF5 data, we use the \mesc\, test suite to pass in an analytic form of the metric, extrinsic curvature, and matter content. This analytic metric is then passed through the regular, general \mesc\, routines which produce output as usual. We can then directly compare the output effective parameters with the analytic solutions to follow.

This test is not intended to place error bars on the results presented in the main text (however, see Appendix~\ref{appx:errors}). Here, we are isolating the \mesc\ calculations from the NR simulations, and not only ensuring that we have small errors, but that these calculations are accurate in reproducing a known analytic solution.

We test our calculations using a linearly-perturbed EdS metric specified by the metric \eqref{eq:longmetric}, which (to linear order in $\phi$) has extrinsic curvature 
%following constraints} \ahr{Do you also want to keep $\beta^i = 0$ here?}
%\begin{subequations}\label{eqs:pertmetric}
\begin{equation}\label{eq:linKij}
    %\alp &= a \sqrt{1+2\phi}, \\
    %\beta^i &= 0, \\
    %\gam_{ij} &= a^2 (1 - 2\phi) \delta_{ij}, \\
    K_{ij} = - a' (1-3\phi) \delta_{ij},
\end{equation}
%\end{subequations}
where $a$ is the EdS scale factor and the operator $'$ represents a derivative with respect to EdS conformal time $\eta$. The metric perturbation, $\phi$, is required to have small amplitude, i.e., $|\phi|\ll 1$.
%We note that for this test, 
The coordinate time satisfies $t=\eta$ for this test 
%-- compare (\ref{eq:longmetric}) with (\ref{eq:ds}) -- 
due to our choice of lapse function.  % in (\ref{eqs:pertmetric}). 
Solving the perturbed Einstein equations for the above metric, we find that the density contrast and the velocity perturbation of the fluid are given by
\begin{align}
    \delta &= - \frac{3}{2} \left( Q a_{\rm init} \xi\right)^2 \, \nabla^2\phi - 2 \phi,\label{eq:lindelta}\\
    v^i &= \frac{Q}{\xi} \delta^{ij} \pd_j \phi,\label{eq:linvel}
\end{align}
with $\delta \equiv \rho/\bar{\rho}-1$, where $\bar{\rho}$ is the EdS background density. %\ahr{I don't think that we use overbar anywhere else? So just said directly what $\bar{\rho}$ is here.} % an overbar represents a background FLRW quantity, 
The constant $Q\equiv - 1 / \sqrt{6\pi\rho^* a_{\rm init}}$ is specified by the background rest-mass $\rho^*\equiv\bar{\rho}a^3$  
and the initial value of the scale factor, $a_{\rm init}$. The operator $\nabla^2 \equiv \delta^{ij} \partial_i \partial_j$ is the spatial Laplacian, and the scaled conformal time is $$\xi\equiv 1 + \sqrt{\frac{2\pi\rho^*}{3a_{\rm init}}}\eta.$$ 
We note that the metric \eqref{eq:longmetric} and extrinsic curvature \eqref{eq:linKij} along with the density perturbation \eqref{eq:lindelta} and the velocity \eqref{eq:linvel} are the equations used to specify initial data in \flrwsolver\ \citep{Macpherson:2016ict,Macpherson:2019a}.

%To simplify the analytic calculations, while still testing all relevant parts of the code, we neglect the acceleration $a^\mu$ \emph{for this test only}. Since the  --- neglecting the acceleration here will have no effect on the results of this test.  
%\ahr{This is something that we know, so erased the 'we expect' part} 
The 4--acceleration is subdominant in this test model setup,
%(which we also find for the simulations presented here), 
and therefore we neglect $a^\mu$ for the case of this test only. This will have no effect on the validity of the test, since derivatives are calculated using finite differences of $\mathfrak{H}$ --- with no explicit reference to the specific form of $\mathfrak{H}$ itself. Therefore, for this section only, the effective Hubble parameter \eqref{def:Eevolution} takes the form $\mathfrak{H} = \frac{1}{3}\theta + e^\mu e^\nu \sigma_{\mu\nu}.$ 
%\ahr{Here I put the formula in the text (referring to the general formula for h), since it is not referred to, and I think it is better to not give it a too prominent position.}

%\begin{equation}
%    \mathfrak{H} = \frac{1}{3}\theta + e^\mu e^\nu \sigma_{\mu\nu}.
%\end{equation}

To first order in all perturbations (and their derivatives), the expansion scalar and shear tensor are, respectively,
\begin{align}
	\theta &= \frac{3a'}{a^2}(1-\phi) + \frac{Q}{\xi}\nabla^2\phi, \\
	\sigma_{ij} &= \frac{Q a^2}{\xi} \left( \pd_i \pd_j \phi - \frac{1}{3}  \nabla^2 \phi \, \delta_{ij} \right),
\end{align}
with $\sigma_{00}=\sigma_{0i}=0$, and $\delta_{ij}$ is the Kronecker delta. The first derivatives of $\mathfrak{H}$ appearing in \eqref{eq:dhds} and \eqref{eq:d2hds2} are
\begin{align}
    \pd_\eta \mathfrak{H} &= \left[ \frac{a''}{a^2} - \frac{2(a')^2}{a^3} \right] (1-\phi) + \frac{1}{9 a} \nabla^2 \phi \label{eq:dth}\\
    &- e^i e^j \sigma_{ij} \left(2\mathcal{H} + \frac{1}{Q a_{\rm ini}\xi} \right), \nonumber\\
    \pd_i \mathfrak{H} &= - \frac{a'}{a^2} \pd_i \phi + \frac{Q}{3\xi} \pd_i \left(\nabla^2\phi\right) + 2 e^k e^l \sigma_{kl} \pd_i \phi \label{eq:dih}\\
    &+ \frac{Q a^2}{\xi} \left[ e^k e^l \pd_i \pd_k \pd_l \phi - \frac{1}{3} \delta_{kl} e^k e^l \pd_i (\nabla^2\phi) \right],\nonumber
\end{align}
where $\mathcal{H}=a'/a$ is the conformal Hubble parameter. The second time derivative of $\Eu$ is
\begin{align}\label{eq:d2th}
    \pd_\eta^2 \mathfrak{H} &= 6 \mathcal{H} \left( \frac{\mathcal{H}^2}{a} - \frac{a''}{a^2}\right) (1-\phi) - \frac{a'}{9 a^2}\nabla^2\phi \\ 
    &- e^i e^j \sigma_{ij} \left[ 2 \mathcal{H}' + \frac{1}{3 (Q a_{\rm ini} \xi)^2} - \left(2\mathcal{H} + \frac{1}{Q a_{\rm ini} \xi}\right)^2 \right],\nonumber
\end{align}
its time-space cross derivative is
\begin{align}\label{eq:didth}
    \pd_i \pd_\eta \mathfrak{H} &= -\pd_i (\phi) \Bigg[ \frac{a''}{a^2} - \frac{2(a')^2}{a^3} \Bigg] + \frac{1}{9a} \pd_i (\nabla^2 \phi) \\
    &- \frac{Q a^2}{\xi} e^k e^l \left[ \pd_i \pd_k \pd_l \phi - \frac{1}{3}\pd_i (\nabla^2 \phi)\delta_{kl} \right]\\
    &\times \left(2\mathcal{H} + \frac{1}{Q a_{\rm ini} \xi}\right),\nonumber
\end{align}
and its second spatial derivative is
\begin{equation}\label{eq:didjh}
    \begin{aligned}
    \pd_i \pd_j \mathfrak{H} &= \pd_i \pd_j \phi \left(2 e^k e^l \sigma_{kl} -\frac{a'}{a^2} \right) \\
    & + \frac{Q}{3\xi} \pd_i \pd_j (\nabla^2 \phi) \left( 1 - a^2 e^k e^l \delta_{kl} \right) \\
	 & + \frac{Qa^2}{\xi} e^k e^l \bigg[ 4 \pd_i (\phi) \pd_j \pd_k \pd_l (\phi) \\
	 &- \frac{4}{3} \pd_i (\phi) \pd_j (\nabla^2\phi) \delta_{kl} + \pd_i \pd_j \pd_k \pd_l (\phi) \bigg].
    \end{aligned}
\end{equation}
To linear order, the components of the Ricci tensor are
\begin{align}
	R_{00} &= - 3\mathcal{H}' + \nabla^2\phi, \\
	R_{0i} &= 2\mathcal{H} \pd_i \phi, \\
	R_{ij} &= \left[ \left(\mathcal{H}^2 + \frac{a''}{a}\right) (1-4\phi) + \nabla^2\phi \right]\delta_{ij},
\end{align}
and the spatial Christoffel symbols are 
\begin{equation}\label{eq:lin_christoffels}
\hspace*{-0.18cm}    \Gamma^i_{jk} = \frac{1}{(1-2\phi)} \! \left[\delta^{il}\pd_l(\phi) \delta_{jk} -\pd_j (\phi) \delta^i_k - \pd_k (\phi) \delta^i_j \right].
\end{equation}
We choose a single mode form of the perturbation $\phi$, namely
\begin{equation}
	\phi = \phi_0 \sum_i {\rm sin}\left(\frac{2\pi x^i}{L}\right),
\end{equation}
where $L=1$ is the length of the test domain, and we set $\phi_0=10^{-8}$ to ensure that higher-order contributions remain below the level of numerical errors.

We use the above expressions to calculate the analytic form of the first \eqref{eq:dhds} and second \eqref{eq:d2hds2} derivatives of $\Eu$ along the null ray. We then use these, along with the analytic Ricci tensor components above, to calculate the analytic solutions for the effective cosmological parameters \eqref{eq:paramseff}.

We run \mesc\, for 1000 randomly placed observers each with 300 randomly chosen lines of sight. For each observer, and for each individual line of sight, we calculate the 
effective cosmological parameters \eqref{eq:paramseff}
%terms in the series expansion \eqref{eq:dLexpand2} 
using the completely general \mesc\, routines, given the analytic metric \eqref{eq:longmetric} and extrinsic curvature \eqref{eq:linKij}, and compare to the analytic expressions shown above. 
We define the relative error, e.g. for the Hubble parameter, as
\begin{equation}\label{eq:err_analytic}
    {\rm err}(\Eu) \equiv \frac{\Eu_o}{\Eu_{o, {\rm analytic}}} - 1,
\end{equation}
which we calculate along all 300 line of sights for each observer.

% ==========================================
% plots of effective params errors and convergence
% ==========================================
\begin{figure*}%[h]
    \centering
    \includegraphics[width=\textwidth]{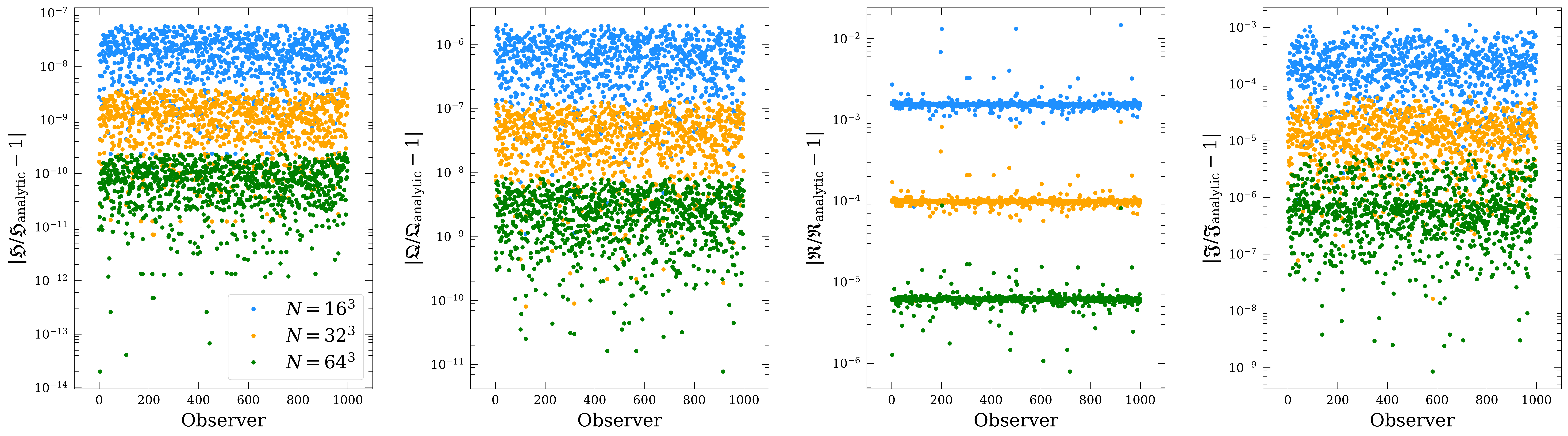}
    \caption{Errors for the effective Hubble, deceleration, curvature, and jerk parameters (panels; left-to-right, respectively), for the test case of an analytic, linearly-perturbed EdS metric passed into \mesc. Each point shows the relative error for an individual observer, averaged over 300 lines of sight. Colours represent different numerical resolutions, as indicated in the legend.
    }
    \label{fig:tests_error}
\end{figure*}

Figure~\ref{fig:tests_error} shows the absolute value of the relative error for the effective Hubble, deceleration, curvature, and jerk parameters (panels, left-to-right, respectively). Points show \eqref{eq:err_analytic} averaged over all lines of sight for a single observer. Blue points are for a $16^3$ grid, yellow for a $32^3$ grid, and green for a $64^3$ grid. The relative error remains below 1\% for the effective curvature parameter, below 0.1\% for the effective jerk parameter, and below $10^{-4}$\% for the effective Hubble and deceleration parameters, even for the coarse resolutions used here. 

The errors in these calculations should reduce at a rate defined by the order of accuracy of the scheme used when increasing the resolution. The rate of convergence, $C$, for a set of errors at three resolutions is calculated as, e.g. for the Hubble parameter,
\begin{equation}\label{eq:conv}
    C(\Eu) \equiv \frac{{\rm err(\Eu)_{low}} - {\rm err(\Eu)_{mid}}}{{\rm err(\Eu)_{mid}} - {\rm err(\Eu)_{high}}},
\end{equation}
where a subscript `low' refers to the error for the lowest resolution, `mid' for the middle resolution, and `high' for the highest resolution. For fourth-order accurate derivatives, and when doubling the resolution for each increase, the expected rate of convergence is $C_{\rm exp}=16$. 
\begin{figure*}%[h]
    \centering
    \includegraphics[width=\textwidth]{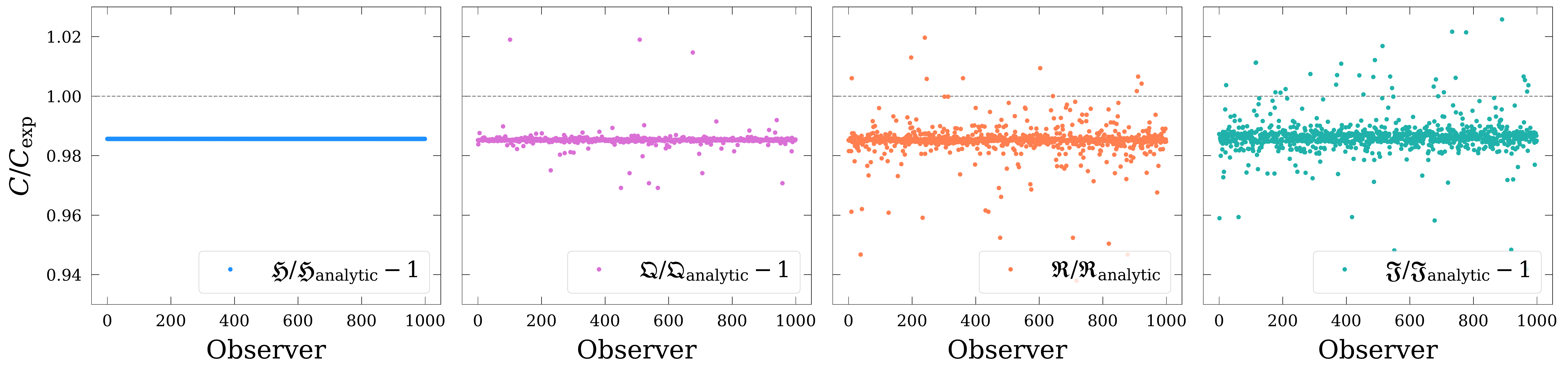}
    \caption{Convergence rate \eqref{eq:conv} relative to the expected value, $C/C_{\rm exp}$, for the effective Hubble, deceleration, curvature, and jerk parameters (panels, left-to-right, respectively). The convergence here is for the errors shown in Figure~\ref{fig:tests_error} at resolutions $16^3, 32^3$, and $64^3$.
    }
    \label{fig:tests_conv}
\end{figure*}
Figure~\ref{fig:tests_conv} shows the convergence rate \eqref{eq:conv}, relative to the expected value, for the errors in the effective cosmological parameters shown in Figure~\ref{fig:tests_error}. All parameters show the expected rate of convergence for fourth-order accurate calculations.

%\hmr{When you change this to params: note that the points that don't converge properly are the ones which have really tiny error at the level of the perturbation, i.e. 1e-8}

%%%%%%%%%%%%%%%%%%%%%%%%%%%%%%%%%%%%%%%%%%%%%%%%%%%%%%%%%%%%%%%%%%%%%%%%%%%%%%%%%%%
\section{Numerical convergence and error bars}\label{appx:errors}
%%%%%%%%%%%%%%%%%%%%%%%%%%%%%%%%%%%%%%%%%%%%%%%%%%%%%%%%%%%%%%%%%%%%%%%%%%%%%%%%%%%

In the previous section we assessed the level of error in the \mesc\ calculations alone, excluding any additional numerical error associated with performing NR simulations. Our main results will contain additional sources of error to those addressed in the previous section. Here we compare simulations at several resolutions to assess convergence and quantify the error bars on our main results.

%\hmr{Do we need this paragraph here? Or are they not different enough for anyone to notice...?}
{We note that the simulations used in this Appendix contain an error in the initial data \citet[see][for details]{MH21:erratum}, however this does not change the numerical convergence properties of the simulations. Fixing this error changes the power spectrum of the initial data, with no effect on the evolution of the simulations and therefore convergence properties will remain unchanged. For this reason we did not consider it necessary to re-run these convergence tests. However, we note this here because the variances presented in these controlled-mode simulations are larger than the main results, due to their higher typical density contrasts.}

The simulations presented here use the Einstein Toolkit paired with \flrwsolver\ for initial data. Our numerical setup is the same as presented in \citet{Macpherson:2019a}, except that we use different power spectra 
%\aht{are} used 
for the initial conditions. 
%Here, we use initial power spectra from CLASS \citep[rather than CAMB \footnote{https://camb.info} as in][]{Macpherson:2019a}, because of the large domain sizes required for the scales we are interested in. To study individual grid cells of hundreds of Mpc in size, we require simulation domains that are 10's of Gpc in size \hmt{to ensure sufficiently high numerical resolution.} CAMB outputs the matter power spectrum in the co-moving synchronous gauge, which is not the gauge used for initial data in \flrwsolver. For the scales studied in \citet{Macpherson:2019a}, the power spectrum is the same in co-moving synchronous gauge and longitudinal gauge. However, in this work we must take the gauge transformation into account, and CLASS allows us to generate power spectra in the longitudinal gauge. 
We use the BSSNOK formalism to evolve our cosmological space-times, which means the constraint equations are not explicitly enforced as a part of this evolution. The level of violation in the constraints therefore tells us how closely our simulations are matching a solution of Einstein's equations. Any violation arises either as a result of constraint violation in the initial data, or numerical error produced during the simulation. We assume linear perturbations in our initial conditions, and therefore we have an initial violation of size second-order in the perturbations. We expect the violation due to numerical error will dominate by the end of the simulation. We use \mesc\ to calculate the $L_1$ norm of the dimensionless Hamiltonian constraint violation, which for the $N=128$ simulations with 100 (200) $h^{-1}$ Mpc smoothing lengths we find to be $0.6\%$ ($1.1\%$). We refer the reader to \citet{Macpherson:2019a} for further details on the constraint violation and studies on its convergence in a set of simulations using the same numerical methods as used here.

\subsection{Controlled-mode simulations}\label{appx:pkcutsims}

We perform simulations at 3 different numerical resolutions to calculate error bars for our main results via a Richardson extrapolation.  %An important point is that  
The Richardson extrapolation method requires keeping the 
%initial conditions of the 
physical system in question fixed with the change of resolution, such that the error associated with the numerical resolution can be isolated.
%The main simulations used in this work are thus not suitable for Richardson extrapolation, since the physical system considered changes with resolution in these simulations.
%We wish to place error bars on our calculations of the parameters \eqref{eq:paramseff}, which involve derivatives of the fluid at a particular observer location. We must ensure that these derivatives, i.e. the local structure, remain constant with any change in numerical resolution to be able to make a meaningful comparison.
The simulation shown in Figure~\ref{fig:2Dslices}, and any others quoted in the main text, represent a ``full'' power spectrum sampling, i.e. the initial data contains modes down to the minimum possible wavelength (equal to the physical scale spanned by 2 grid cells). As we increase numerical resolution (reduce the size of grid cells) we therefore change the \emph{physical} local structure at each observer's position. These simulations are thus not suitable for a Richardson extrapolation. 

Therefore, as in \citep{Macpherson:2019a}, we perform a set of simulations in which small-scale structure is removed from the initial data, and only long-wavelength modes are sampled initially \citep[see also][]{Giblin:2016mjp,Daverio:2019gql}. If these modes are sufficiently large, we expect the structures at all points in the domain to remain similar between resolutions. 
We perform three simulations with domain length 2 $h^{-1}$ Gpc at resolution $N^3$ for $N=32,64$, and $128$, in which we have excluded any power at scales below $10 \Delta_{x, 32}$, where $\Delta_{x, 32}$ is the grid spacing for the $N=32$ simulation. Higher resolution initial data is obtained by interpolating the lower-resolution data. The minimum modes therefore have wavelength $\sim 625 h^{-1}$ Mpc in all initial data. %Importantly, we note that these are strictly the scales sampled in the initial data \emph{only}, and due to the inherently nonlinear form of the simulations, there is no guarantee that smaller-scale structures will not appear at late times in the simulation.
\begin{figure}%[h]
    \centering
    \includegraphics[width=\columnwidth]{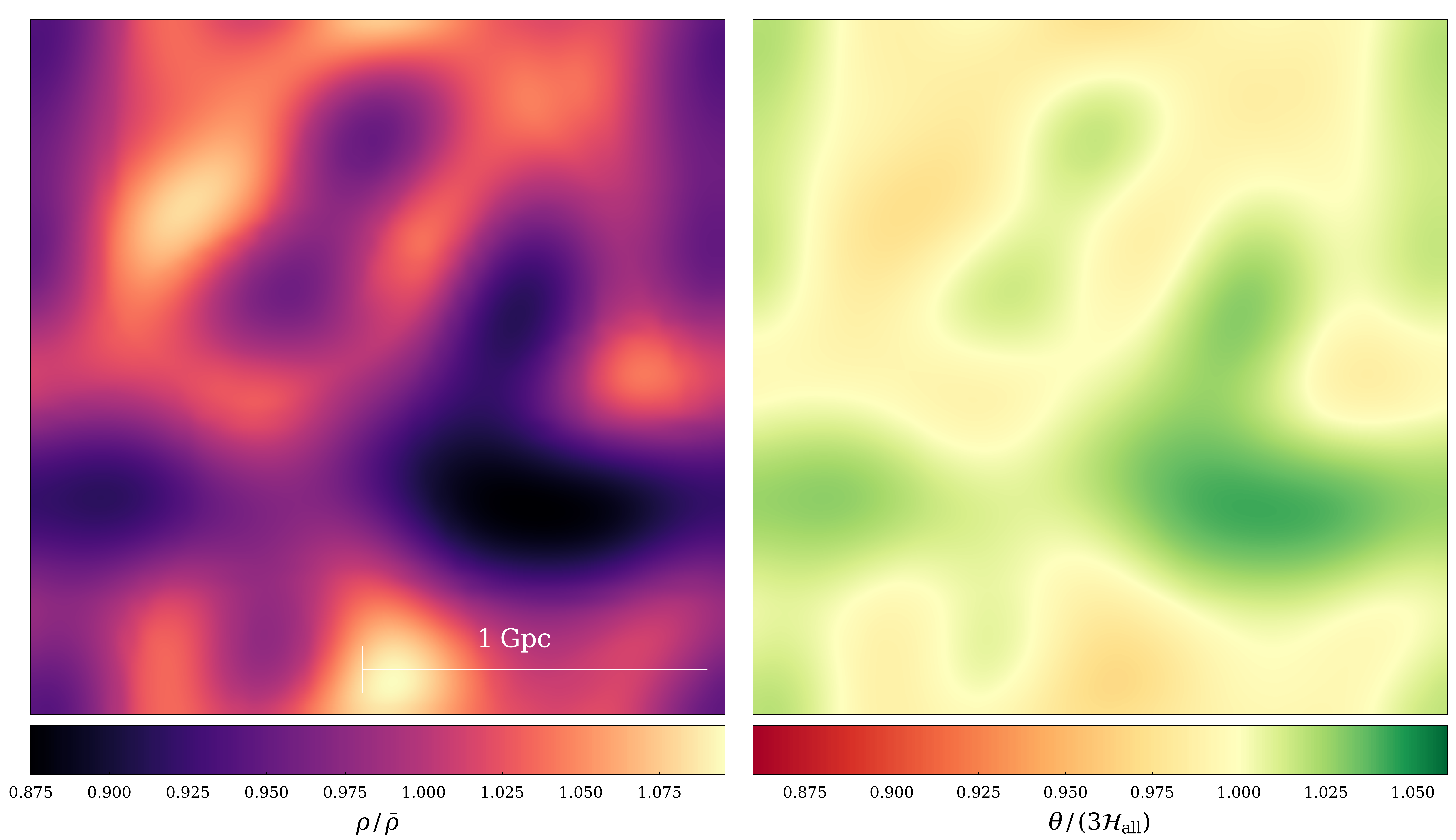}
    \caption{Density (left panel) and expansion rate (right panel) for our $N=64$ test simulation used to perform a Richardson extrapolation of our results. We show a 2--dimensional slice through the domain at $\zeff=0$, for a $2 h^{-1} {\rm Gpc}$ domain length. 
    }
    \label{fig:convsim}
\end{figure}
As an example, Figure~\ref{fig:convsim} shows a 2--dimensional slice through the $N=64$ test simulation, with the density field, $\rho$, in the left panel and the expansion scalar, $\theta$, in the right panel. 

One might consider using simulations such as this for the main results presented in this paper. 
%However, we use the full power spectrum sampled simulations for the following reasons. 
While we do restrict the minimum mode to wavelength $\sim 625 h^{-1}$ Mpc for the initial data for the simulations described in this appendix, there is no guarantee that structures below this scale will not form later in the simulation. In this work, we wish to strictly constrain any structures to be above the smoothing scale of interest, which is why we choose individual grid cells to coincide with this scale. However, we confirm that the effective parameters are qualitatively similar in the controlled-mode simulations when sampling similar physical scales.

\subsection{Richardson extrapolation and definition of error}\label{appx:richextrap}

We expect that our numerical estimates of physical quantities of interest will approach the ``true'' values of the quantities (in the context of the approximations and limitations of the simulations used) as we increase numerical resolution towards infinity, i.e., $N\rightarrow\infty$. The rate at which an estimated quantity will approach its ``true'' value depends on the accuracy of the implemented scheme (i.e., how the error reduces as we increase resolution). The Einstein Toolkit thorns which we use are fourth-order accurate (as is \mesc), and we therefore expect our numerical estimates to converge at a rate $\propto N^{-4}$. Thus, we estimate the numerical error of a quantity by calculating the same quantity at three different resolutions, and fitting a curve of the form $f(N) = a + b/N^4$, where $a$ and $b$ are parameters determined using the \texttt{SciPy}\footnote{\url{https://scipy.org}} package's \texttt{curve\_fit} function. Extrapolating the quantity to very large $N$ by using the determined function $N \mapsto f(N)$ (here we choose $N=10^5$, which ensures that the value is stable) gives an estimate of the ``true'' value of that quantity. We then define the error in, e.g., $\Eu$ as the residual between our highest-resolution calculation and the extrapolated ``true'' value, normalised by the mean magnitude of $\Eu$, namely,
\begin{equation}\label{eq:errordef}
    \Eu \; {\rm error} \equiv \frac{\Eu(N=128) - \Eu_{\rm extrap}(N=10^5)}{{\rm mean}(|\Eu(N=128)|)}.
\end{equation}
Normalising the error in this way produces an estimate on the relative error while avoiding spurious large-magnitude errors caused by near-zero values (especially relevant for those parameters with distributions centred around zero).

In the controlled simulations described in the previous section, we calculate the effective cosmological parameters \eqref{eq:paramseff} at 1000 observer positions, keeping these positions constant between resolutions. For each observer, we randomly choose 300 lines of sight across the whole observer's sky (`FullSky' in Figure~\ref{fig:fullsky}). The direction vector $e^\mu$ for each line of sight, for each observer, is determined using the same random numbers 
%\ahr{changed to $m^i$ here, correct?} 
$m^i$ (see Appendix~\ref{sec:emu}) between resolutions. However, in order to ensure the orthonormality requirements (\ref{eq:emuconstraints}), we require the 3--metric and fluid 4--velocity at each location to determine $e^\mu$ from $m^i$. Therefore, the direction vectors do vary \emph{slightly} between resolutions, however, we confirm that the components of $e^\mu$ generally converge at the expected fourth-order rate. 
%\ahr{Precision question: Is it the direction vectors themselves that converge at this order, or quantities derived from them?}\hmr{the components of $e^\mu$ converge, I guess this is not surprising since the metric and 4-velocity converge, but still maybe worth stating?} \ahr{yes I agree!}

\begin{figure*}%[h]
    \centering
    \includegraphics[width=0.8\textwidth]{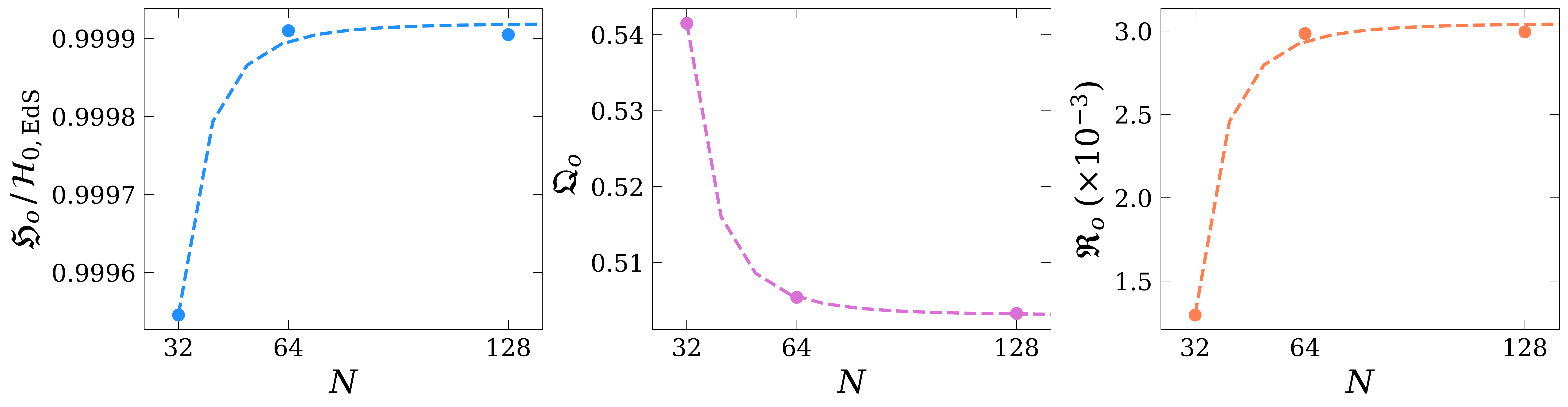}
    \caption{Example of the Richardson extrapolation process for effective Hubble, deceleration, and curvature parameters (panels, left-to-right, respectively) for controlled-mode simulations with resolutions $N=32,64,$ and $128$ (x-axis). Points are calculations from the simulations averaged over 1000 observers, and dashed curves are the best-fit curves representing fourth-order convergence.
    }
    \label{fig:richextrap}
\end{figure*}
We average the effective anisotropic cosmological parameters of each observer over all 300 lines of sight, and compare this average between resolutions using a Richardson extrapolation. We also assess the convergence of the parameters averaged over all 1000 observers. Figure~\ref{fig:richextrap} shows an example of this process for the effective Hubble, deceleration, and curvature parameters (panels, left-to-right, respectively, see Appendix~\ref{sec:jerkerror} below for a discussion on the error in the jerk parameter) for the average over all 1000 observers. Points show calculations from simulations at resolution $N$ (x-axis) and dashed curves are the best-fit representing fourth-order convergence.
The asymptotic values in Figure~\ref{fig:richextrap} can be interpreted as reflecting the expected ``true'' values for each parameter (for a particular coarse-graining scale), i.e. at infinite numerical resolution. We note that we do not necessarily expect these asymptotic values to correspond to their EdS counterparts, since the points themselves represent different numerical resolutions, and not different coarse-graining scales. We can interpret each point in Figure~\ref{fig:richextrap} in the same way as the horizontal lines in, e.g. Figure~\ref{fig:200_params_full} (or Figure~\ref{fig:200_params_allres}), i.e. the average over all observers.
We further stress that the non-zero asymptotic value of the effective curvature parameter, in the right-most panel of Figure~\ref{fig:richextrap}, is not necessarily indicative of a non-zero averaged spatial curvature in the simulation. This is due to the fact the interpretation of the generalised curvature parameter is not simply connected to the averaged Ricci curvature of spatial sections, as it is in the FLRW cosmography {\citep[see][]{Heinesen:2020b}}. 

%\ahr{I replaced the citation \citep{Heinesen:2020a} with \citep{Heinesen:2020b} here, since I think that it best describes the interpreation of the effective curvature parameter here. (even though discussions on curvature in \citep{Heinesen:2020a} are also interesting :) )}

\subsection{Error calculation in controlled-mode simulations}

%From the process outlined in the previous section, we estimate the error using  for each observer.
\begin{figure*}%[h]
    \centering
    \includegraphics[width=0.8\textwidth]{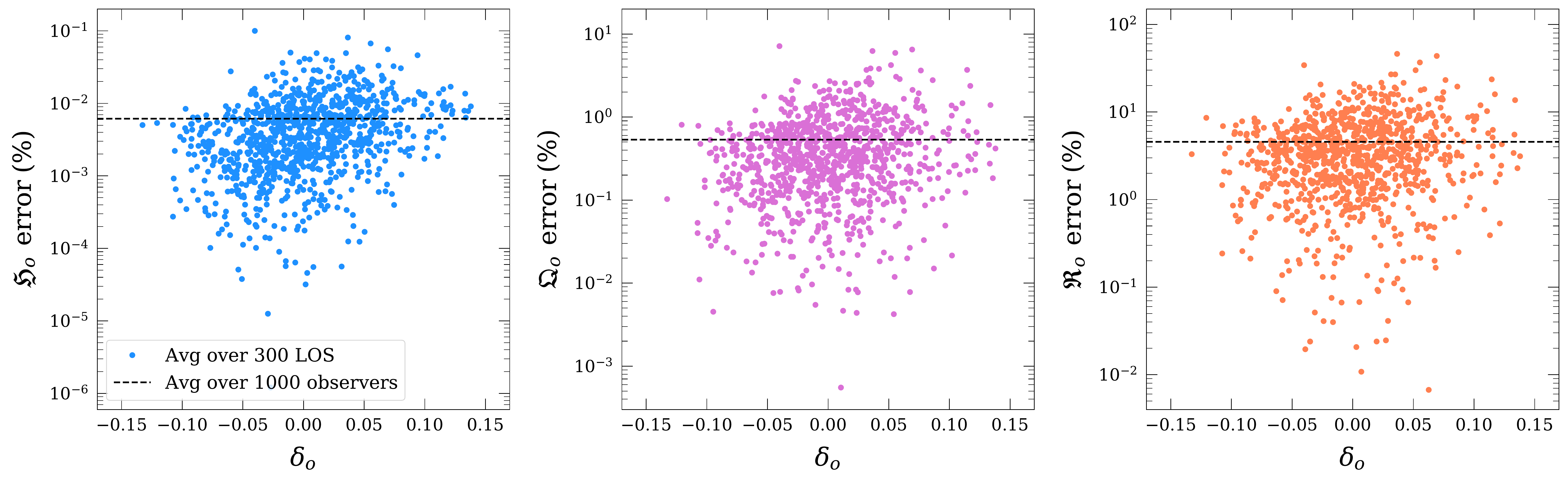}
    \caption{Percentage relative error \eqref{eq:errordef} for effective Hubble, deceleration, and jerk parameters (panels, left-to-right) calculated via a Richardson extrapolation. Points show the error in calculations averaged over 300 lines of sight for each observer, and black dashed lines in each panel show the average error over all 1000 observers.
    }
    \label{fig:paramerrs}
\end{figure*}
Panels, left-to-right, in Figure~\ref{fig:paramerrs} show the error \eqref{eq:errordef} for the effective Hubble, deceleration, and curvature parameters, respectively, as a function of the local over-density at the observer, $\delta_o$. Points show the error in an observers line-of-sight average measurement, and dashed black lines show the average error over all observers, which yield $\sim 0.006\%$, $\sim 0.5\%$, and $\sim 4.6\%$ for the effective Hubble, deceleration, and curvature parameters, respectively. See Section~\ref{sec:jerkerror} below for a discussion on the error in the jerk parameter. The controlled-mode simulations used to quantify these errors are sampling similar physical scales at $\zeff=0$ (of the order of hundreds of Mpc), and so we take the errors shown in Figure~\ref{fig:paramerrs} to be representative of the level of error in our main results.

As mentioned above, there is no guarantee that smaller-scale structures will not develop in these controlled-mode simulations. This also implies that differences in structure growth between resolutions are possible. In turn, not all observers will necessarily have similar enough local environments, and in these cases we expect that calculations will not converge. This can simply be because differences between resolutions are no longer purely due to truncation error in the finite difference approximation of derivatives. 

This appears to be the cause of the largest-magnitude errors in all parameters shown in Figure~\ref{fig:paramerrs}. Specifically, we find that for observers with the largest errors --- mainly coinciding with non-convergence of the respective parameter ---, the local density $\delta_o$ also does not converge. This indicates that the structure is physically different between resolutions at the position of these observers and therefore a meaningful comparison between resolutions is difficult.

\subsection{Statistical convergence}

As detailed above, in order to obtain convergence for an \emph{individual} observer we require the local structure at that observer's position to remain constant as we increase resolution. However, as long as the simulations in question continue to sample the same physical smoothing scale, we expect to see \emph{statistically} similar results across all observers, even when increasing numerical resolution. In this section, we ensure that our main results exhibit statistical convergence in this sense.

We perform three simulations each with a coarse-graining scale of 200 $h^{-1}$ Mpc, with resolutions $N=32, 64$, and $128$. The highest resolution of this set is the main simulation presented in Figure~\ref{fig:2Dslices}. Since we must maintain the same minimum scales sampled, with each $2\times$ increase in resolution we also increase the total (physical) domain size by a factor of 2. The simulation box sizes are therefore $6.4, 12.8$, and $25.6 h^{-1}$ Gpc, respectively. The size of the box in code units remains the same for all simulations in this study, and all initial data are different realisations of the same power spectrum. %, ensuring structure on the scale of the observer is statistically consistent. 
\begin{figure}%[h]
    \centering
    \includegraphics[width=\columnwidth]{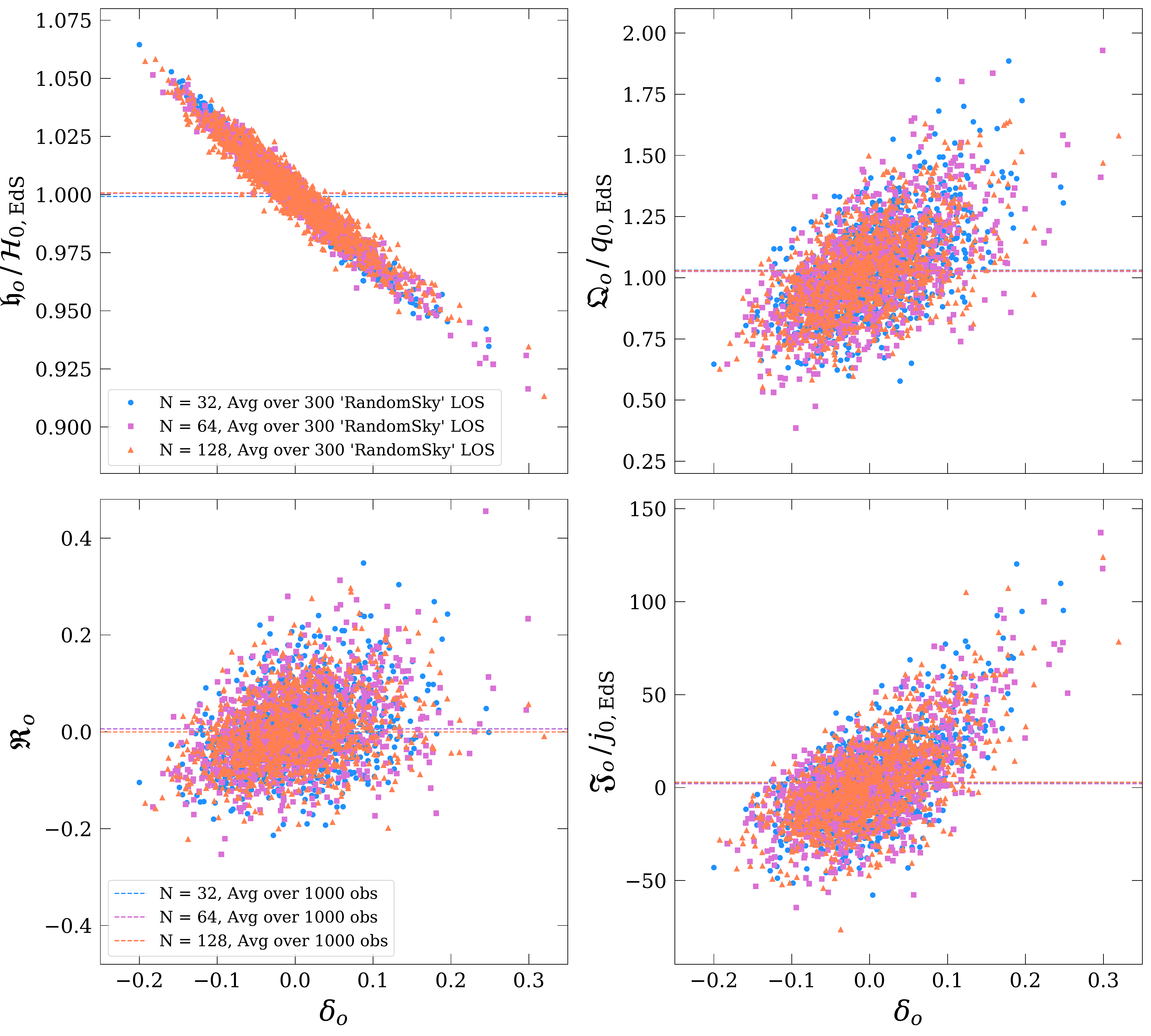}
    \caption{Effective cosmological parameters for 3 simulations at different numerical resolution (different coloured points and lines), all with the same physical smoothing length of 200 $h^{-1}$ Mpc. This figure shows the same as Figure~\ref{fig:200_params_full} for multiple resolutions sampling the same physical scale.
    }
    \label{fig:200_params_allres}
\end{figure}

Points in Figure~\ref{fig:200_params_allres} show the effective cosmological parameters \eqref{eq:paramseff} as measured by 1000 observers (each averaged over 300 `FullSky' lines of sight) in these three simulations (different colours, as indicated in the legend). Dashed lines of the same colour show the average over all observers for each resolution. Qualitatively, we notice the distribution in each parameter does not change drastically with resolution. We perform a Kolmogorov-Smirnov (KS) test to assess the null hypothesis that parameter distributions between resolutions $N=64$ and $128$ are drawn from the same distribution. We find that the KS test fails to reject the null hypothesis at significance level $\alp=0.15$. We therefore conclude that the results presented in the main text regarding statistics across all observers are robust to changes in resolution. 
%\hmr{note the minimum p-value is about 0.1}

\subsection{Error in the jerk parameter}\label{sec:jerkerror}

For the controlled-mode simulations discussed in the previous sections, we did not find the expected convergence of the effective jerk parameter, $\mathfrak{J}_o$. We found that for $\sim 15\%$ of observers, $\mathfrak{J}_o$ is well-behaved and converges at the expected fourth-order rate. For these observers the maximum relative error in $\mathfrak{J}_o$ is $\sim 5\%$. However, for most observers, and therefore for the average over all observers, we do not see convergence.

We have attributed this to differences in the local 
%\ahr{density gradients?} 
density gradients at late times, which, as discussed in the previous sections, can be slightly different between resolutions. We confirm that the jerk parameter converges at the expected fourth-order rate at earlier times in the simulations, when gradients are more similar between resolutions. We also refer the reader to Section~\ref{sec:analytic_test}, where $\mathfrak{J}_o$ converges as expected in an analytic test where gradients are constructed to be exactly identical between resolutions. We note that for the observers who do not show convergence at late times, we also do not see convergence of the local density, $\delta_o$, whereas for those 15\% who do show convergence, the local density also converges. This strongly suggests the main reason for this is due to differences in the local density field. 
The dominant contribution to the jerk parameter is the second derivative of $\Eu$ along the null ray (i.e., \emph{third} derivatives of the fluid velocity). If local structure is slightly different between resolutions this will be more noticeable in the calculation of higher-order derivatives. This explains both why we see convergence in $\mathfrak{J}_o$ at earlier times, and why we see convergence of other parameters (which have dominant contributions from lower-order derivatives) at late times. 

Since we cannot quantify error bars on $\mathfrak{J}_o$ in a controlled simulation that is still physically similar to our main results, we choose to avoid making statements on the jerk parameter for individual observers. From the results of the KS test presented in the previous section, the \emph{distribution} of the jerk parameter across all observers exhibits convergence. %We therefore are confident in the statistical properties of our calculation of the jerk parameter. 

\bibliography{refs}{}

\end{document}